\pdfoutput=1

\RequirePackage{etoolbox}
\csdef{input@path}{%
 {sty/}% cls, sty files
 {img/}% eps files
}%
\csgdef{bibdir}{bib/}% bst, bib files

\documentclass[ba]{imsart}
\pubyear{2019}
\volume{00}
\issue{0}
\doi{0000}
\firstpage{1}
\lastpage{1}

\usepackage{amsthm}
\usepackage{amsmath}
\usepackage{natbib}
\usepackage[colorlinks,citecolor=blue,urlcolor=blue,filecolor=blue,backref=page]{hyperref}
\usepackage{graphicx}
\usepackage{amsfonts}
\usepackage[outdir=./]{epstopdf}
\usepackage{subcaption} 
\usepackage{bm}
\usepackage{bbm}
\usepackage{placeins}
\usepackage{natbib}
\usepackage{algorithm}
\usepackage{algpseudocode}
\theoremstyle{plain}
\newtheorem{theorem}{Theorem}

\newtheorem{algo}{Algorithm}
\newtheorem{fact}{Fact}

\newcommand{\Normal}[2][\bm{0}]{\mathcal{N}(#1, #2)}
\newcommand{\btheta}{\bm{\theta}}
\newcommand{\momentum}{\bm{p}}
\newcommand{\chainState}{\bm{z}}
\newcommand{\Mass}{\bm{M}}
\newcommand{\Fe}{F_{\epsilon}}

\newcommand{\given}{\, | \,}

\newcommand{\T}{\mathcal{T}}
\newcommand{\A}{\mathcal{A}}

\algrenewcommand\algorithmicindent{1.0em}

\startlocaldefs
\numberwithin{equation}{section}
\theoremstyle{plain}

\endlocaldefs

\begin{document}

%% *** Frontmatter *** 

\begin{frontmatter}
	\title{Recycling Intermediate Steps to Improve Hamiltonian Monte Carlo}
	\runtitle{Recycled Hamiltonian Monte Carlo}
%	\thankstext{T1}{Footnote to the title with the ``thankstext'' command.}
	
	\begin{aug}
		\author{\fnms{Akihiko} \snm{Nishimura}\thanksref{addr1}\ead[label=e1]{akihiko4@ucla.edu}}
		\and
		\author{\fnms{David} \snm{Dunson}\thanksref{addr2}\ead[label=e2]{dunson@duke.edu}}
		
		\runauthor{Nishimura and Dunson}
		
		\address[addr1]{Department of Biomathematics, University of California Los Angeles, Los Angeles, CA 90095, USA
			\printead{e1} 
		}
		
		\address[addr2]{Department of Statistical Science, Duke University, Durham, NC 27708, USA
			\printead{e2}
		}
		
	\end{aug}
	
	\begin{abstract}
	Hamiltonian Monte Carlo (HMC) and related algorithms have become routinely used in Bayesian computation. In this article, we present a simple and provably accurate method to improve the efficiency of HMC and related algorithms with essentially no extra computational cost.  This is achieved by {\em recycling} the intermediate states along simulated trajectories of Hamiltonian dynamics. Standard algorithms use only the end points of trajectories, wastefully discarding all the intermediate states.  Compared to the alternative methods for utilizing the intermediate states, our algorithm is simpler to apply in practice and requires little programming effort beyond the usual implementations of HMC and related algorithms. Our algorithm applies straightforwardly to the no-U-turn sampler, arguably the most popular variant of HMC. Through a variety of experiments, we demonstrate that our recycling algorithm yields substantial computational efficiency gains.
	\end{abstract}
	
	
\end{frontmatter}

\section{Introduction}
\label{sec:intro}
Markov chain Monte Carlo is routinely used to generate samples from posterior distributions. While specialized algorithms exist for restricted model classes, general-purpose algorithms are often inefficient and scale poorly in the number of parameters. Originally proposed by  \citet{duane87} and popularized in the statistics community through the works of \citet{neal96, neal10}, Hamiltonian Monte Carlo promises a better scalability \citep{neal10, beskos13} and has enjoyed wide-ranging successes as one of the most reliable approaches in general settings \citep{bda13, kruschke14, monnahan16}.  Stan and PyMC software packages take advantage of this generality and performance \citep{stan15, pymc16}.

Given a parameter $\btheta \sim \pi_{\btheta}(\cdot)$ of interest, HMC introduces an auxiliary \textit{momentum} variable $\momentum$ and defines a distribution $\pi(\cdot) = \pi_{\btheta}(\cdot) \times \Normal{\Mass}$ on the augmented parameter space $(\btheta, \momentum)$ with a \textit{mass matrix} $\Mass$. A proposal is generated by simulating trajectories of \textit{Hamiltonian dynamics} where the evolution of the state $(\btheta, \momentum)$ is governed by a differential equation:
\begin{equation} 
\label{eq:Hamilton}
% \left\{
\begin{aligned}
\frac{{\rm d} \btheta}{{\rm d} t}
&= \Mass^{-1} \momentum, \quad 
\frac{{\rm d} \momentum}{{\rm d} t}
= \nabla \log \pi_{\btheta}(\btheta).
\end{aligned}
% \right.
\end{equation}
Proposals generated by this mechanism can be far away from the current state and yet accepted with high probability. In case $\Mass = \bm{I}$ and $\btheta \in \mathbb{R}^2$, the solution trajectory of \eqref{eq:Hamilton} coincides with the motion of a frictionless puck sliding over a surface of height $-\log \pi(\btheta)$ \citep{neal10}. The puck feels ``push'' in the direction of the gradient $\nabla \log \pi(\btheta)$, pointing toward higher values of $\log \pi(\btheta)$. The momentum evolves accordingly but also gives the puck tendency to keep moving in the same direction, helping HMC proposals explore the parameter space in an informed manner.

% This behavior is due to the following property of \eqref{eq:Hamilton}: if $\{ (\btheta(t),\momentum(t)) \}_t$ denotes the solution of the differential equation with the initial condition $(\btheta(0), \momentum(0)) = ( \btheta_0, \momentum_0 ) \sim \pi(\cdot)$, then $(\btheta(t), \momentum(t)) \sim \pi(\cdot)$ for all $t \in \mathbb{R}$. 
In practice, the analytical solution to \eqref{eq:Hamilton} is rarely available and a trajectory $( \btheta(t), \momentum(t) )$ for $0 \leq t \leq \tau$ is approximated by taking $K \approx \tau / \epsilon$ steps of a \textit{leap-frog} scheme with stepsize $\epsilon$, where each step 
$\Fe: (\btheta_0, \momentum_0) \to (\btheta_1, \momentum_1)$ is defined via the relations
\begin{equation}
\label{eq:leapfrog}
\begin{aligned}
\momentum_{1/2} - \momentum_0
&= \frac{\epsilon}{2} \nabla \log \pi_{\btheta}(\btheta_0) \\
\btheta_1 - \btheta_0
&= \epsilon \, \Mass^{-1} \momentum_{1/2} \\
\momentum_1 - \momentum_{1/2}
&= \frac{\epsilon}{2} \nabla \log \pi_{\btheta}(\btheta_1).
\end{aligned}
\end{equation}
% $$ \momentum_{1/2} - \momentum_0 = \frac{\epsilon}{2} \nabla \log \pi_{\btheta}(\btheta_0), \quad \btheta_1 - \btheta_0 = \epsilon \, \momentum_{1/2}, \quad \momentum_1 - \momentum_{1/2} = \frac{\epsilon}{2} \nabla \log \pi_{\btheta}(\btheta_1). $
The approximate solution $\Fe^K(\btheta_0, \momentum_0) \approx ( \btheta(\tau), \momentum(\tau) )$ no longer has the distribution $\pi(\cdot)$, but can be used as a Metropolis proposal \citep{metropolis53}. Efficiency of HMC depends critically on a choice of $\epsilon$, $\tau$, and $\Mass$, but well-established approaches exist for tuning these parameters \citep{andrieu08, neal10, wang13, hoffman14, stan15}.

Current practice uses the last state $\Fe^K(\btheta_0, \momentum_0)$ as a proposal and discards all the intermediate values $\Fe^k(\btheta_0, \momentum_0)$ for $k < K$. As we will show, this is wasteful since the intermediate values can be \textit{recycled} to generate additional samples from posterior distributions. Figure~\ref{fig:recycling_illustration}, to be explained in detail later, illustrates the benefit of recycling. The recycling algorithm only requires quantities that have already been sampled or computed, so there is essentially no extra computational cost. Our proposed recycling approach can also be applied directly to a wide variety of modified HMC algorithms \citep{neal10, girolami11, pakman13, pakman14, lan14, shahbaba14, fang14, zhang16, lu16}. Extensions to more complex variants are also possible, including the No-U-Turn-Sampler (NUTS) \citep{hoffman14, stan15}. 

Our algorithm is distinguished by its simplicity and generality compared to alternative algorithms for utilizing the intermediate values of HMC \citep{neal94, calderhead14, bernton15}. Under our framework, one can typically implement an HMC variant as usual and simply add several lines of code to recycle the intermediate values using the familiar acceptance and rejection probabilities. 
On the other hand, the algorithm of \cite{calderhead14} and its Rao-Blackwellization by \cite{bernton15} require a trajectory to be simulated forward and backward in a symmetric manner to satisfy the \textit{super-detailed balance} condition \citep{frenkel04, tjelmeland04}. The proposal must then be followed by the acceptance-rejection step using the generalized Metropolis-Hastings algorithm \citep{calderhead14} or assignment of appropriate weights to the intermediate values \citep{bernton15}. Importantly, these algorithms do not apply to NUTS, arguably the most popular variant of HMC. This is because NUTS yields a variable number of intermediate states and does not constitute a multi-proposal scheme necessary for using their algorithms.

The underlying idea behind our algorithm is most similar to \cite{neal94} who realized that, in the variant of HMC that uses a collection of states in computing the acceptance probability, those states can be re-used when computing the posterior summaries through conditional expectation. Our theory is more general, however, and translates into methods to improve a variety of multi-proposal algorithms (Section~\ref{sec:theory}). 
% Also justified by our theory are various schemes to select only a subset of the intermediate states to recycle; this is an important feature in a high-dimensional parameter space where the memory requirement to store the extra samples becomes substantial (see Section~\ref{sec:simulation}). 

\section{Recycled Hamiltonian Monte Carlo}
\label{sec:main_idea}
The following non-standard HMC algorithm accepts or rejects each of the intermediate values, enabling recycling of these samples. 
%The number of steps $L^{(i)}$ is randomized as recommended in the literature to avoid periodic behavior in the trajectories of \eqref{eq:Hamilton} \citep{neal10}.
\begin{algo}[Recycled HMC] 
	\label{alg:rehmc}
	Generate random variables $\{ (\btheta_k^{(i)}, \momentum_k^{(i)}),$ $k = 0,1,\ldots,K \}_{i \geq 1}$ so that the sequence 
	$\{ (\btheta_0^{(i)}, \momentum_0^{(i)}) \}_{i \geq 1}$ forms a Markov chain with transition rule $(\btheta_0^{(i)}, \momentum_0^{(i)}) \to (\btheta_0^{(i+1)}, \momentum_0^{(i+1)})$ as follows:
	\begin{enumerate}
		\item Set $\btheta_0 = \btheta_0^{(i)}$ and draw a random momentum $\momentum_0 \sim \Normal{\Mass}$.
		\item Simulate a trajectory via the leapfrog steps as in \eqref{eq:leapfrog} and set $(\btheta_k^*, \momentum_k^*) = \Fe^k(\btheta_0, \momentum_0)$ for $k = 1, \ldots, K$.
		\item  Accept or reject each state along the trajectory;  for $k = 1, \ldots, K$, set $(\btheta_k^{(i+1)}, \momentum_k^{(i+1)}) = (\btheta_k^*, \momentum_k^*)$ with probability
		\begin{equation}
		\label{eq:HMC_ar_step}
		\min \left\{1, 
		\frac{
			\pi\big( \btheta_k^*, \momentum_k^* \big)
		}{
			\pi \big( \btheta_0, \momentum_0 \big)
		} \right\} 
		\end{equation}
		and $(\btheta_k^{(i+1)}, \momentum_k^{(i+1)}) = (\btheta_0, \momentum_0)$ otherwise. 
		\item Update the starting point of the next HMC iteration: $(\btheta_0^{(i+1)}, \momentum_0^{(i+1)}) = (\btheta_{K}^{(i)}, \momentum_{K}^{(i)})$.
		%		\begin{equation}
		%		(\btheta_k^{(i)}, \momentum_k^{(i)}) = 
		%		\left\{
		%		\begin{array}{ll}
		%		\Fe^k(\btheta_0^{(i)}, \momentum_0^{(i)}) 
		%			& \text{with probability} \
		%		\min \left\{1, 
		%			\frac{ \pi\left( \Fe^k(\btheta_0^{(i)}, \momentum_0^{(i)}) \right)
		%			}{
		%			\pi \left( (\btheta_0^{(i)}, \momentum_0^{(i)}) \right) } \right\} \\
		%			(\btheta_0^{(i)}, \momentum_0^{(i)}) & \text{ otherwise}
		%		\end{array}
		%		\right.
		%		\end{equation}
%		\item Set $(\btheta_0^{(i+1)}, \momentum_0^{(i+1)}) = (\btheta_{L^{(i)}}^{(i)}, \momentum_{L^{(i)}}^{(i)})$ for $L^{(i)}$ drawn from a distribution $\pi_L(\cdot)$ on $\{1, \ldots, K \}.$
	\end{enumerate}
\end{algo}

The transition rule $(\btheta_0^{(i)}, \momentum_0^{(i)}) \to (\btheta_0^{(i+1)}, \momentum_0^{(i+1)})$ above coincides with that of the standard HMC algorithm. 
The empirical measure $N^{-1} \sum_{i=1}^N  \delta_{(\btheta_0^{(i)}, \momentum_0^{(i)})} (\cdot)$ thus converges weakly to the target distribution.
The following theorem, which is a consequence of a more general theory given in the next section, shows that the intermediate samples can actually be recycled as valid draws from the target distribution.
The index $k$ runs from 1 to $K$ as $(\btheta_0^{(i+1)}, \momentum_0^{(i+1)}) = (\btheta_{K}^{(i)}, \momentum_{K}^{(i)})$ and $k = 0$ is redundant.
\begin{theorem}
	\label{thm:reHMC}
	If $(\btheta_k^{(i)}, \momentum_k^{(i)})$ for $k = 1, \ldots, K$ are generated as in Algorithm \ref{alg:rehmc}, then
	\begin{equation} 
	\frac{1}{N K}\sum_{i=1}^N \sum_{k =1}^{K} \delta_{(\btheta_k^{(i)}, \momentum_k^{(i)})} (\cdot) \overset{w}{\to} \pi(\cdot)
	\ \text{ as } \ N \to \infty,
	\end{equation}
	where $\overset{w}{\to}$ denotes the weak convergence of a measure.
\end{theorem}
%The proof follows from a more general theory given in the next section.

\begin{figure}
	\hspace{-3ex}
	\centering
	\includegraphics[width=.65\textwidth]{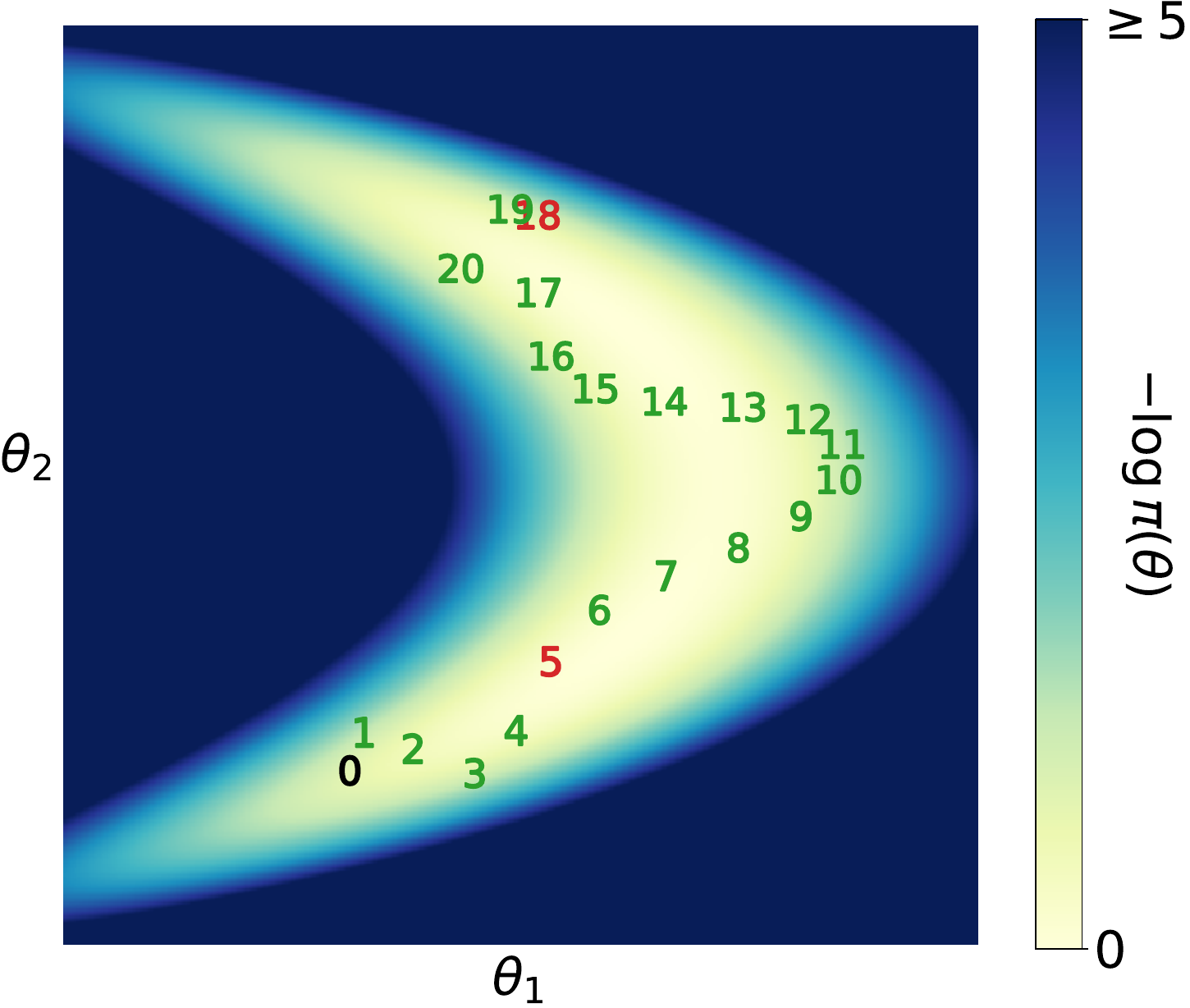} 
	\caption{Visual demonstration of the recycling algorithm applied during one iteration of HMC on a 2-D banana-shaped target distribution. The numbers $k = 0, 1, \ldots, 20$ indicate the location $(\btheta_k, \momentum_k)$ of a simulated trajectory after $k$ leapfrog steps. Red color at $k = 5$ and $k= 18$ indicates rejection of the intermediate states. Green color indicate the successfully recycled (i.e.\ accepted) intermediate states. Most of the intermediate states have high acceptance rates as expected from the property of Hamiltonian dynamics.
%	The target distribution is obtained as a posterior distribution under a model $y \given x_1, x_2 \sim \normal(\textrm{mean} = x_1 + x_2^2, \textrm{sd} = 0.4), x_1, x_2 \sim \normal(0, 1)$. The trajectory was generated with the initial condition $q_0 = (-.75, -1.25), \momentum_0 = (0.1, 1.0)$ with $\epsilon = .2$.
	}
	\label{fig:recycling_illustration}
\end{figure}

Fig.~\ref{fig:recycling_illustration} illustrates the recycling procedure. The Metropolis type acceptance-rejection of \eqref{eq:HMC_ar_step} is applied to each intermediate state $\Fe^k(\btheta_0^{(i)}, \momentum_0^{(i)})$. Calculating these acceptance probabilities typically takes little additional computational time; the unnormalized target densities at the intermediate values are either already computed in common variants of HMC \citep{neal10, hoffman14} or can be obtained cheaply as a by-product of computing the gradients $\nabla \log \pi_{\btheta}$.

Algorithm~\ref{alg:rehmc} uses the fixed path length $K$, but it is often desirable to randomize the number of steps to avoid periodic behavior in the trajectories of \eqref{eq:Hamilton} \citep{neal10}. Recycling is justified under this setting as well:
\begin{theorem}
	\label{thm:general_ReHMC}
	Consider a modified version of Algorithm~\ref{alg:rehmc} in which at the $i$-iteration 1) the trajectory is simulated for a random path length $L^{(i)} \sim \pi_L(\cdot)$ with $L^{(i)} \leq K$ and 2) the starting point for the next HMC iteration is set as $(\btheta_0^{(i+1)}, \momentum_0^{(i+1)}) = (\btheta_{L^{(i)}}^{(i)}, \momentum_{L^{(i)}}^{(i)})$. Then we have
	\begin{equation}
	\frac{1}{\sum_{i=1}^N L^{(i)}} \sum_{i=1}^N \sum_{k =1}^{L^{(i)}} \delta_{(\btheta_k^{(i)}, \momentum_k^{(i)})} (\cdot) \overset{w}{\to} \pi(\cdot)
	\text{ as } N \to \infty.
	\end{equation}
	%	Moreover, if $\btheta_0 \sim \pi(\cdot)$ and $K_i = K$ for all $i$, then
	%		\begin{equation}
	%		\Var( \frac{1}{M} \sum_{i=1}^M g(\btheta_L^{(i)}) )
	%			> \Var( \frac{1}{M K} \sum_{i=1}^M \sum_{k =1}^K \g(\btheta_k^{(i)}) )
	%		\end{equation}
\end{theorem}

\section{Theory Behind Recycling Algorithm}
\label{sec:theory}

The validity of recycled HMC as in Theorem~\ref{thm:reHMC} and \ref{thm:general_ReHMC} follows from a more general principle below. 

%The principal can be applied in the same manner to a range of HMC extensions \citep{neal10, girolami11, fang14}, recycling all the intermediate states of a trajectory by simply adding acceptance-rejection steps. The application to a more complex proposal generation mechanism is also possible minor modifications and will be illustrated with NUTS in the next section.

\begin{theorem}
	\label{thm:vector_ergodic_theorem}
	Let $P_k(\cdot \given \cdot)$ for $k = 0, \ldots, K$ be transition kernels with a common stationary measure $\pi(\cdot)$ and suppose $P_0(\cdot \given \cdot)$ is uniquely ergodic.\footnote{A transition kernel (or a Markov chain) with a unique stationary measure is called \textit{uniquely ergodic}. The uniqueness of a stationary measure implies ergodicity by the ergodic decomposition theorem \citep{kallenberg02}.}
	%$\su\momentum_{\chainState_0} \Vert P_0^n(\cdot \given \chainState_0) - \pi(\cdot) \Vert_{\rm tv} \to 0$ as $n \to \infty$ where $P_0^n(\cdot \given \chainState_0)$ denotes the $n$-step transition probability and $\Vert \cdot \Vert_{\rm tv}$ a total variation distance.
	Consider a Markov chain $\{\chainState^{(i)}\}_{i \geq 1}$ on a product space $\chainState = (\chainState_0, \ldots, \chainState_{K})$ whose transition probability $\chainState \to \chainState^*$ only depends on the coordinate $\chainState_0$ i.e.\
	\begin{equation} 
	\label{eq:assump_first_coord_only}
	P(\chainState^*_0, \ldots, \chainState^*_K \given \chainState_0, \ldots, \chainState_K)
	= P(\chainState^*_0, \ldots, \chainState^*_K \given \chainState_0)
	\end{equation}	
	and has the marginal densities
	\begin{equation}
	\label{eq:assump_marginal}
	\int P(\chainState^*_0, \ldots, \chainState^*_K \given \chainState_0) \, {\rm d} \chainState^*_{-k}
	= P_k(\chainState^*_k \given \chainState_0)
	\end{equation}	
	where $\chainState^*_{-k} = (\chainState^*_0, \ldots, \chainState^*_{k-1}, \chainState^*_{k+1}, \ldots, \chainState^*_K)$ for $k = 0, \ldots, K$. Then the following result holds:
	\begin{equation}
	\label{eq:vector_ergodic_theorem}
	\frac{1}{N K}\sum_{i=1}^N \sum_{k =1}^{K} \delta_{\chainState_k^{(i)}} (\cdot) \overset{w}{\to} \pi(\cdot)
	\ \text{ as } \ N \to \infty.
	\end{equation}
	Additionally, the Markov chain $\{\chainState^{(i)}\}_{i \geq 1}$ is geometrically (or uniformly) ergodic if $P_0(\cdot \given \cdot)$ is geometrically (or uniformly) ergodic.
\end{theorem}

\begin{proof} %[Theorem~\ref{thm:vector_ergodic_theorem}]
	It is easy to verify that the Markov chain $\chainState^{(1)}, \chainState^{(2)}, \ldots$ has a stationary distribution 
	\vspace{-.8ex}
	\begin{equation}
	\vspace{-.6ex}
	\pi^*(\cdot) = \int P(\, \cdot \given \chainState_0) \pi(\chainState_0) \, {\rm d} \chainState_0.
	\end{equation}
	By the assumption \eqref{eq:assump_marginal}, the marginal $\pi^*(\chainState_k)$ coincides with $\pi(\chainState_k)$ for all $k = 0, \ldots, K$. Once we establish the unique ergodicity of the chain $\{ {\chainState}^{(i)} \}_{i \geq 1}$, therefore, the conclusion \eqref{eq:vector_ergodic_theorem} follows by averaging the coordinates $\chainState_1, \ldots, \chainState_k$ of the empirical measure $N^{-1} \sum_{i=1}^N \delta_{\chainState^{(i)}}$. 
	
	To see that $\pi^*(\cdot)$ is the unique stationary measure, suppose $\tilde{\pi}^*(\cdot)$ is another stationary measure of $P(\cdot \given \cdot)$. This means that, by the assumption \eqref{eq:assump_first_coord_only},
	\vspace{-.8ex}
	\begin{equation}
	\vspace{-.5ex}
	\label{eq:another_stat_meas}
	\tilde{\pi}^*(\cdot)
	= \int P(\, \cdot \given \chainState_0) \tilde{\pi}^*(\chainState_0) \, {\rm d} \chainState_0.
	\end{equation}	
	In particular, the marginal $\tilde{\pi}^*(\chainState_0^*)$ satisfies $\tilde{\pi}^*(\chainState_0^*) = \int P_0(\chainState_0^* \given \chainState_0) \tilde{\pi}^*(\chainState_0) \, {\rm d} \chainState_0$ by the assumption  \eqref{eq:assump_marginal}. The unique ergodicity of $P_0(\cdot \given \cdot)$ then implies $\tilde{\pi}^*(\chainState_0) = \pi(\chainState_0)$. Substituting this equality into \eqref{eq:another_stat_meas} establishes $\tilde{\pi}^*(\cdot) = \pi^*(\cdot)$ and hence the unique ergodicity of the chain $\{ {\chainState}^{(i)} \}_{i \geq 1}$.
	
	We turn to the proof of a convergence rate (geometric or uniform ergodicity) of the chain $\{ {\chainState}^{(i)} \}_{i \geq 1}$ under the corresponding assumption on $P_0(\cdot \given \cdot)$. For the conditional distribution of $\chainState^{(n)} \given \chainState_0^{(0)}$, we have
	\begin{equation}
	\begin{aligned}
	\left| \mathbb{P}(\chainState^{(n)} \in A \given \chainState_0^{(0)}) - \pi^*(A) \right|
	&= \left| \int P(A \given \chainState_0') \left( P_0^n(\chainState_0' \given \chainState_0^{(n)}) - \pi(\chainState_0') \right) \mathrm{d} \chainState_0' \right|. \\
	%&\leq \left\| P_0^n(\, \cdot \given \chainState_0^{(n)}) - \pi(\cdot) \right\|_{\mathrm{tv}}
	\end{aligned}
	\end{equation}
	It follows that 
	\begin{equation*}
	\left\| \mathbb{P}(\chainState^{(n)} \in \cdot \given \chainState_0^{(0)}) - \pi^*(\cdot) \right\|_{\mathrm{tv}}
	\leq \left\| P_0^n(\, \cdot \given \chainState_0^{(0)}) - \pi(\cdot) \right\|_{\mathrm{tv}}
	\end{equation*}	 
	where $\| \cdot \|_{\mathrm{tv}}$ denotes a total variation norm. Hence the chain $\{ {\chainState}^{(i)} \}_{i \geq 1}$ inherits the convergence rate of $P_0(\cdot \given \cdot)$.
\end{proof}

Theorem~\ref{thm:vector_ergodic_theorem} has a subtle but important difference from ``composition sampling,'' in which one would first generate a Markov chain $\{\chainState_0^{(i)}\}_{i \geq 0}$ and then sample $(\chainState_1^{(i+1)}, \ldots, \chainState_K^{(i+1)})$ from a conditional distribution $\pi^*(\, \cdot \given \chainState_0^{(i)})$. For a Markov chain generated as in Theorem~\ref{thm:vector_ergodic_theorem}, the conditional distribution $\chainState_1^{(i+1)}, \ldots, \chainState_K^{(i+1)} \given \chainState_0^{(i)}$ may have dependency on $\chainState_0^{(i+1)}$. This additional flexibility is critical for the recycling algorithms presented in this article. 
In particular, Theorem~\ref{thm:vector_ergodic_theorem} reduces to Theorem~\ref{thm:reHMC} when the transition kernel $P_k(\cdot \given \cdot)$ is constructed as one iteration of HMC with $k$ leapfrog steps for $k \geq 1$ and $P_0(\cdot \given \cdot)$ as that with $K$ steps.

Theorem~\ref{thm:general_ReHMC} is justified by an extension of Theorem~\ref{thm:vector_ergodic_theorem}, which shows that a recycling algorithm applies even when the number of states generated by a multi-proposal scheme varies from one iteration to another. Since  in Theorem~\ref{thm:general_ReHMC} the number of recyclable samples varies according to the random path length $L^{(i)}$'s, the general theory behind Theorem~\ref{thm:general_ReHMC} relies on the framework of a Markov chain in a trans-dimensional parameter space \citep{hastie2012reversible-jump}. The precise statement is given as Theorem~\ref{thm:variable_size_vector_ergodic_theorem} along with a proof in Appendix~\ref{app:recycling_variable_number_of_states}. 

The general formulation of the recycling algorithm as in Theorem~\ref{thm:vector_ergodic_theorem} and \ref{thm:variable_size_vector_ergodic_theorem} is of practical value for any MCMC algorithm that simultaneously yields multiple valid transition kernels $P_k(\cdot \given \cdot)$'s. Indeed, in many variants of HMC \citep{neal10, girolami11, shahbaba14, fang14}, a proposal is generated by computing a long trajectory whose intermediate steps constitute valid proposal states that can be all recycled by simply adding acceptance-rejection steps as in Algorithm~\ref{alg:rehmc}. Our theory also provides an alternate and simpler justification of the algorithms by \cite{calderhead14} and \cite{bernton15} as shown in Appendix~\ref{app:simple_proof_of_calderhead}. Our recycling algorithm can also be applied to more complex proposal generation mechanisms as we illustrate in the next section. 

%In fact, we will develop an improved recycling algorithm for NUTS based on another generalization of Theorem~\ref{thm:vector_ergodic_theorem}, relaxing the assumption that the states $\chainState_0^{(i+1)}, \ldots, \chainState_K^{(i+1)}$ conditionally on $\chainState_0^{(i)}$ are generated independently.	

\section{Recycled No-U-Turn-Sampler}
\label{sec:recycled_NUTS}
No-U-turn sampler (NUTS) of \cite{hoffman14} automates choice of path lengths by simulating each trajectory of Hamiltonian dynamics until it starts moving back towards the starting point, a criteria they termed the \textit{U-turn} condition. The length of a trajectory is recursively doubled forward or backward in a randomly chosen direction. This generates a trajectory of length $2^d$ endowed with a binary tree structure, where $d$ denotes the depth at which the U-turn condition occurs.
% NUTS ensures reversibility of the chain by checking the U-turn condition for the entire tree as well as all its subtrees.

Unlike the simpler trajectory simulation procedure behind HMC, the trajectory doubling procedure of NUTS does not yield a sequence of valid intermediate proposals. In particular, the empirical distribution does not converge to the correct target distribution if we naively recycle all the intermediate states of NUTS as in Algorithm~\ref{alg:rehmc}. A simple recycling algorithm for NUTS can nonetheless be devised by taking advantage of the fact below. Instead of the Metropolis acceptance-rejection procedure, NUTS determines acceptable states along a simulated trajectory using a slice sampling approach via an auxiliary slicing variable $u$.
% TODO: cite a slice sampling paper.
%Note that the transition kernel $P_1(\, \cdot \given \cdot \,)$ below is different from that of NUTS:
\begin{fact}
	\label{fact:nuts_recycle_uniformly}
	The following transition rule $(\btheta_0, \momentum_0) \to (\btheta^*, \momentum^*)$ preserves the target distribution $\pi(\cdot)$. Let $\mathcal{T} = \mathcal{T}(\btheta_0, \momentum_0)$ denote a collection of $2^d$ states generated by one NUTS iteration from the initial state $(\btheta_0, \momentum_0)$, including $(\btheta_0, \momentum_0)$ itself. Generate $u \sim \text{Unif}\left([0, \pi(\btheta_0, \momentum_0)]\right)$ and sample $(\btheta^*, \momentum^*)$ uniformly from the collection of acceptable states
	\begin{equation}
	\label{eq:nuts_acceptable_states}
	\A = \A \left(\T, u\right)
	= \left\{ (\btheta, \momentum) \in \T \given \pi(\btheta, \momentum) > u \right\}.
	\end{equation}
	% (The construction of $\A$ using the slicing variable $u$ corresponds to the acceptance-rejection step in HMC.)
\end{fact}
\vspace{-\parskip}
\noindent The algorithmic details behind construction of the set $\A$ are complex. Since it is not essential for understanding our recycling approach, we refer the readers to \cite{hoffman14}. The discussions there also justify the above transition rule.

Fact~\ref{fact:nuts_recycle_uniformly} motivates the following algorithm for utilizing the intermediate states generated during each iteration of NUTS. 
% We will describe a more statistically efficient version Algorithm~\ref{alg:ReNUTS} momentarily. 
\begin{algo}[Simple Recycled NUTS]
	\label{alg:simple_ReNUTS}
	Run NUTS to generate a sequence of random variables $\{ (\btheta_0^{(i)}, \momentum_0^{(i)}) \}_{i \geq 1}$. Additionally at each iteration of NUTS, generate  $\{ (\btheta_k^{(i)}, \momentum_k^{(i)}), k = 1, \ldots, K \}$ by sampling $K$ variables without replacement from the acceptable states $\A \big(\T(\btheta_0^{(i-1)}, \momentum_0^{(i-1)}) \big)$ as in \eqref{eq:nuts_acceptable_states}. 
	%Return $\{ (\btheta_k^{(i)}, \momentum_k^{(i)}), k = 1, \ldots, K \}_{i \geq 1}$ as samples from the target distribution.
\end{algo}
\vspace{-\parskip} 
\noindent Fact~\ref{fact:nuts_recycle_uniformly} tells us that the transition $(\btheta_0^{(i)}, \momentum_0^{(i)}) \to (\btheta_k^{(i)}, \momentum_k^{(i)})$ preserves the target distribution $\pi(\cdot)$ for each $k = 1, \ldots, K$. Algorithm~\ref{alg:simple_ReNUTS} is thus justified with a straightforward application of Theorem~\ref{thm:vector_ergodic_theorem}. 

Algorithm~\ref{alg:simple_ReNUTS} captures the main idea behind our recycling algorithm; however, it is actually more statistically efficient to sample $(\btheta_1^{(i)}, \momentum_1^{(i)}), \ldots, (\btheta_K^{(i)}, \momentum_K^{(i)})$ from $\A \big(\T(\btheta_0^{(i-1)}, \momentum_0^{(i-1)}) \big)$ so that they are evenly spread along a NUTS trajectory. Such a sampling scheme can be implemented in a simple and memory efficient manner --- without storing all the intermediate states in memory --- by taking advantage of the binary tree structure of a NUTS trajectory. This is described in Appendix~\ref{app:efficient_recycled_NUTS}

%A statistically efficiency of Algorithm~\ref{alg:simple_ReNUTS} can be improved with a minor modification and this is described in the supplemental appendix.

%Having to choose the number of recycled variables $K$ might be somewhat awkward when the average trajectory length of NUTS for a given target distribution is unknown a priori. In situations where 
When we are not constrained by memory, the following Rao-Blackwellized version of recycled NUTS allows us to simply collect and use all the acceptable states of each NUTS iteration by assigning appropriate weights.
\begin{algo}[Rao-Blackwellized Recycled NUTS]
	\label{alg:rao_blackwell_ReNUTS}
	Denote  the collection of acceptable states from the $i$-th iteration of NUTS by $\A_i = \{ (\btheta_k^{(i)}, \momentum_k^{(i)}), k = 1, \ldots, | \A_i | \}$.
	% Let $\A_i = \{ (\btheta_k^{(i)}, \momentum_k^{(i)}), k = 1, \ldots, | \A_i | \}$ denote the collection of acceptable states from the $i$-th iteration of NUTS. 
	Return the samples $\{ (\btheta_k^{(i)}, \momentum_k^{(i)}), k = 1, \ldots, |\A_i| \}$ with weight $\propto |\A_i|^{-1}$ for $i = 1, \ldots, N$ as the draws from the target distribution, yielding an empirical measure: 
	\begin{equation*}
	\frac{1}{N}\sum_{i=1}^N \frac{1}{|\A_i|} \sum_{k=1}^{|\A_i|} \delta_{(\btheta_k^{(i)}, \momentum_k^{(i)})} (\cdot).
	\end{equation*}
\end{algo}
\vspace{-\parskip}
\noindent The validity of Algorithm~\ref{alg:rao_blackwell_ReNUTS} follows simply by taking an expectation over the sampling step $(\btheta_k^{(i)}, \momentum_k^{(i)}) \sim {\rm Uniform}(\A_i)$ of Algorithm~\ref{alg:simple_ReNUTS}. 

\section{Numerical results}
\label{sec:simulation}

We demonstrate the benefit of recycling using four test cases: three taken from \cite{hoffman14} and one from \cite{girolami11}.
We focus on these well-established test cases as guaranteeing a rapid convergence and mixing of HMC in full generality is an active area of research \citep{livingstone16, livingstone2017kinetic-energy-choice, mangoubi2017hmc-mixing-logconcave}.
In all our simulations we chose the stepsizes $\epsilon$ such that the corresponding average acceptance rates are approximately $70 \%$, as values between $60\%$ and $80\%$ are typically considered optimal \citep{neal10, beskos13}. The dual averaging algorithm of \cite{hoffman14} was used to find such stepsizes. The choice of path lengths $\tau^{(i)} = \epsilon L^{(i)}$ for HMC is discussed within the individual test cases below. The identity mass matrix $\Mass = \bm{I}$ was used in all our simulations except when investigating the use of recycling in mass matrix tuning (see Section~\ref{sec:metrics}).

\subsection{Metrics for computational and memory efficiency of recycling}
\label{sec:metrics}

\subsubsection*{Effective sample sizes for mean, variance, and quantile estimators}
In comparing the algorithms with and without recycling, we use effective sample sizes (ESS) as a commonly used measure of computational efficiency of Monte Carlo algorithms \citep{mcmchandbook11}. 
ESS for a statistics of interest $\mathbb{E}[f(\btheta)]$ is defined via the relation
\begin{equation}
\label{eq:ess_standard_definition}
\textrm{Var}\left(N^{-1} \textstyle \sum_{i=1}^N f(\btheta^{(i)})\right)
	 = {ESS}^{-1} \textrm{Var}\left(f(\btheta^{(1)})\right),
\end{equation}
comparing MCMC samples to the equivalent number of i.i.d.\ samples.
The standard definition above applies only to estimators of the form $N^{-1} \sum_{i=1}^N f(\btheta^{(i)})$ for a real-valued function $f$, so we extend it to a more complex estimator $F: \{\btheta^{(i)}\}_{i=1}^N \to \mathbb{R}$ of a quantity $\mathbb{E}[f(\btheta)]$ by defining
\begin{equation}
\label{eq:ess_definition}
{\rm ESS}_F \left(\{\btheta^{(i)}\}_{i=1}^N \right) = 
N \frac{ 
	{\rm MSE}\left( F \left( \left\{ \btheta^{*(i)} \, \overset{{\rm  i.i.d. }}{\scalebox{1.5}[1]{$\sim$}} \, \pi(\cdot) \right\} \right) \right) 
}{
	{\rm MSE}\left( F(\{\btheta^{(i)}\}) \right)
}
\end{equation}
where MSE$(\cdot)$ denotes the mean squared error of an estimator. The definition \eqref{eq:ess_definition} agrees with the standard one when $F\left(\{\btheta^{(i)}\}_{i=1}^N \right) = N^{-1} \sum_{i=1}^N f(\btheta^{(i)})$. 
% Since additional computer time incurred by recycling is typically insignificant (e.g.\ $1 \sim 6$\% in our examples implemented in \textsc{MATLAB}), the comparison in terms of ESS practically accounts for computational time. % 1% for HBLR and 6% for Gauss & SV

In each of our examples, we compute the ratios of ESS with and without recycling for the mean, variance, and 97.5\% quantile estimator along each model parameter. The ESS ratio is obtained from the ratio of MSE using the relation \eqref{eq:ess_definition}. The MSE of the Monte Carlo estimates are estimated by running 400 chains independently for 3,200 iterations starting from stationarity.\footnote{To obtain stationary samples,  we use the last sample of the long ground-truth NUTS chain as the starting point and run an additional 100 iterations to ensure that the starting points of the 400 chains are independent. We assessed sampling efficiency of NUTS via ESS to ensure that 100 iterations are more than enough to yield an independent sample.}
The number of iterations here is chosen to roughly agree with the typical use cases of HMC --- for example, Stan's default setting generates 1,000 NUTS samples after 1,000 warm-up iterations \citep{stan15}.
We find in many cases, however, that the magnitudes of ESS gains from recycling are independent of the chain lengths (Appendix~\ref{app:ess_ratio_as_function_of_chain_length}).
In computing the MSE of a Monte Carlo estimate, we need to have a precise value for the estimated quantity. Since the analytical expressions are unavailable in our test cases (except for the first synthetic one), we run a long NUTS chain of $10^7$ iterations (after $10^3$ burn-ins) to obtain the ``ground truths'' that are orders of magnitudes more accurate than the individual Monte Carlo estimates from the chains of length 3,200.
% We also extract stationary samples from this long NUTS chain by collecting samples every $25{,}000 = 10^7 / 400$ iterations.
% \footnote{We use the analytical expressions for the synthetic multivariate Gaussian example of Section~\ref{sec:Gaussian_test_case}.}

\subsubsection*{Benefits of recycling in estimating covariance structures}
We also assess whether recycling helps estimate the covariance structure of the target distribution. To this end, we compute the top eigenvalue and eigenvector of the empirical covariance matrix for each chain. We then calculate the angle between the empirical eigenvector and the plane spanned by the $\ell$ true leading principal components. This angle should be close to 0 when the eigenvector is estimated well. To ensure identifiability of the direction, we choose $\ell = \min\{ j: \sigma_j^2 < \sigma_1^2 / 2 \}$ in all our simulations where $\sigma_j^2$ denotes the $j$th largest eigenvalue of the true covariance matrix. 

As an alternate and more holistic evaluation of covariance estimation with a practical application, we further investigate the utility of recycling during the tuning phase of HMC/NUTS.
% during which a covariance matrix of the target $\pi(\cdot)$ is estimated.
Often, initial iterations of HMC/NUTS are used to estimate the covariance matrix of the target $\bm{\Sigma}  = \textrm{Var}(\btheta)$, which can then be used to accelerate HMC/NUTS by setting the mass matrix $\Mass = \widehat{\bm{\Sigma}}^{-1}$ \citep{neal10, stan15}.
If recycling improves the covariance estimator $\widehat{\bm{\Sigma}}$ during the tuning phase, later HMC/NUTS iterations will be faster and mix better.
%Such use of recycling requires no extra memory,
 
To quantify the benefit of recycling in this setting, we estimate the covariance with and without recycling during the tuning phase, and then run two independent chains with the two covariance estimators to compare their ESSs. Recycling is only applied during the tuning phase for covariance estimation. 
In carrying out this experiment, we follow the default settings of Stan for tuning the stepsize and mass matrix.
First, 50 iterations of the dual-averaging algorithm are run to tune the stepsize with the identity mass matrix, followed by $N_{\rm adap}$ iterations with a fixed stepsize to estimate the covariance matrix, and finally another 75 iterations of dual-averaging to re-adjust the stepsize with the tuned mass matrix. After the covariance estimation phase with $N_{\rm adap}$ iterations, we set $\Mass^{-1} = \widehat{\bm{\Sigma}}$ where
%\vspace{-.1ex}
\begin{equation}
\vspace{-.1ex}
\widehat{\bm{\Sigma}} = \frac{N_{\rm adap}}{5 + N_{\rm adap}}\widehat{\bm{\Sigma}}_{\rm emp} + \frac{5}{5 + N_{\rm adap}} 10^{-3} \cdot \bm{I},
\end{equation}
with $\widehat{\bm{\Sigma}}_{\rm emp}$ the empirical covariance matrix and $\bm{I}$ the identity matrix. After the tuning phase, we run NUTS until the total number of gradient evaluations reaches $10^4$. This procedure is repeated 400 times and the ESS for each statistic is averaged across the repetitions. This experiment is not run for the models of Section~\ref{sec:SV_test_case} and \ref{sec:logGaussCox_test_case} as the high-dimensionality of the parameter spaces make covariance estimation impractical.

\subsubsection*{Memory and statistical efficiency trade-off}
We also study the relationship between the number of recycled samples and statistical efficiency. In particular, we demonstrate that it is not necessary to recycle all the intermediate steps to reap the benefit of recycling. This is relevant in a high dimensional parameter space where the amount of memory required to store the extra samples becomes substantial.\footnote{In the stochastic volatility model of Section~\ref{sec:SV_test_case}, for example, it requires 4GB of memory to store 100 extra samples per iteration from a Markov chain of length 3,200 in a 3000-dimensional parameter space.} For long trajectories, there is substantial correlation among the intermediate states and we can expect that recycling a subset of the intermediate states may provide almost as much statistical efficiency as recycling all. 

To quantify this, we first run the algorithm recycling all the intermediate states. We then repeatedly reduce $K$, the number of samples per iteration (recycled intermediate states plus the final state), by a factor of 2. 
In comparing the algorithms with and without recycling, we use the smallest $K$ for which the ESS averaged across all the estimators is within 5\% of that when recycling all the intermediate states. 
Section~\ref{sec:how_many_to_recycle} investigates in more detail the relationship between the statistical efficiency and the number of recycled samples.

\subsection{Multivariate Gaussian}
\label{sec:Gaussian_test_case}

The first test case is sampling from a 250-dimensional multivariate Gaussian $\Normal{{\bm{\Sigma}}}$, where ${\bm{\Sigma}}$ is drawn from a Wishart distribution with 250 degrees of freedom and mean equal to the identity matrix. A covariance matrix drawn from this distribution exhibits strong correlations, and in our case the ratio between the largest and smallest eigenvalue of ${\bm{\Sigma}}$ was approximately $9.5 \times 10^4$. Since HMC and NUTS with the identity mass matrix are invariant under rotations, we take $\Sigma$ to be diagonal with $\Sigma_{i,i} = \sigma_i^2$, where $\sigma_i^2$ corresponds to the $i$th smallest eigenvalue of the original covariance matrix. For the path length of HMC, we first found the smallest value of $\tau_{}$ for which the samples in the leading principal component direction are roughly independent. The typical practice would be then to jitter $\tau^{(i)}$'s within the range $[0.9 \, \tau_{}, 1.1 \, \tau_{}]$ to avoid periodicity \citep{neal10}, but this still resulted in near perfect periodicity and hence poor mixing for some parameters. After some experiments, we found jittering $\tau^{(i)}$ in the range $[\tau / 2, \tau]$ to provide decent mixing along all the coordinates.

\begin{figure}[htb]
	\centering
	\begin{subfigure}[t]{0.45\textwidth}
		\includegraphics[width=\textwidth]{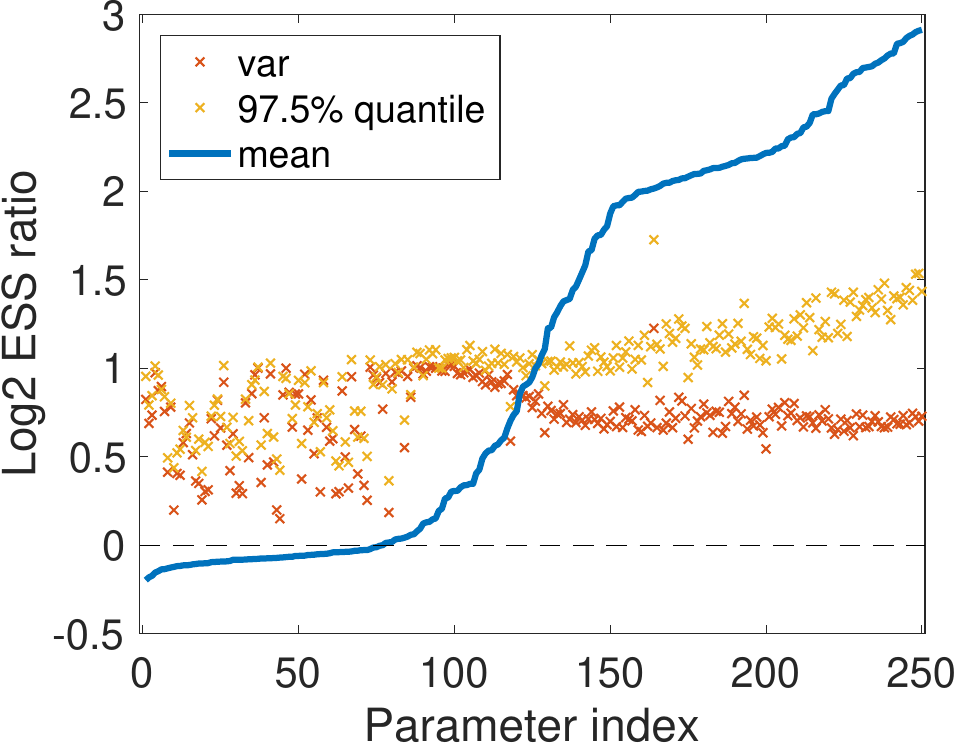}
		\caption{$\log_2$ ratios of ESS with recycling (numerator) and without recycling (denominator). The horizontal line at zero corresponds to no gain from recycling.}
		% The parameter indices ($x$-axis) are sorted by the ESS ratios in mean estimation.
		\label{fig:Gaussian_mean_est}
	\end{subfigure} 
	~
	\begin{subfigure}[t]{0.45\textwidth}
		\centering
		\includegraphics[width=\textwidth]{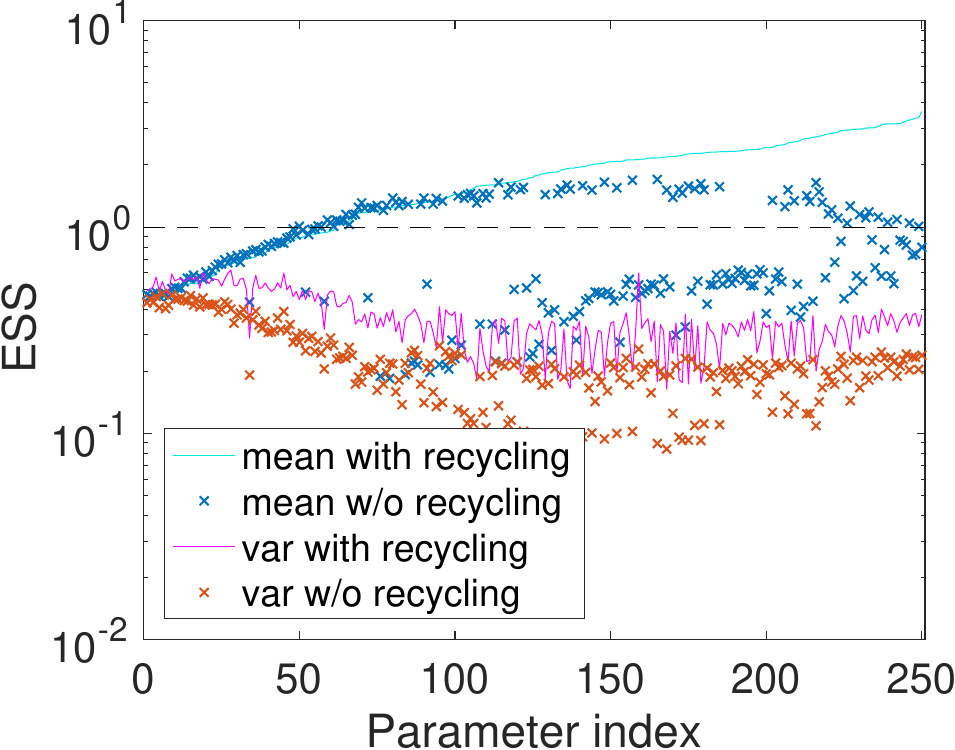}
		\caption{ESS per HMC iteration for the mean and variance estimators. The parameters with the smallest ESS are those with the worst mixings. The y-axis is in $\log_{10}$ scale.} 
			%  i.e.\ this quantity being $10^0 = 1$ means one iteration of (recycled) HMC provides the same amount of information as one independent sample.
		\label{fig:Gaussian_ess_est}
	\end{subfigure}
	\caption{Performance comparison between HMC with and without recycling in estimating mean, variance, and quantiles for the Gaussian example.}
\end{figure}

%We simulated 800 independent Markov chains of length 1600 starting from stationarity. We then computed the MSE in Monte Carlo estimates of the mean, variance, and 97.5\% quantile along each dimension. Fig.~\ref{fig:Gaussian_mean_est} shows $\log_2$ of the ratios between ESS of HMC with and without recycling, calculated from the MSE using the relation \eqref{eq:ess_definition}
Fig.~\ref{fig:Gaussian_mean_est} shows $\log_2$ of the ratios between ESS of HMC with and without recycling. 
To facilitate the comparison of algorithms with and without recycling, the parameters are sorted in increasing order of the ESS ratios in mean estimation. 
Values above zero indicate superior performance of our recycling algorithm. 
The use of recycling improves, uniformly and substantially, estimation of the variances and quantiles: about 100\% increase in ESS on average. 
Not all the mean estimators are improved by recycling, but the ones with worst ESS are significantly improved (Fig.~\ref{fig:Gaussian_ess_est}). 
Out of 251 recyclable samples generated on average from each iteration of HMC, we recycled $251 / 8 \approx 31$ samples.
%\footnote{For this specific example, recycling more samples can substantially boost the ESS of mean estimators for the parameters with small variances. In determining the number of recycled samples through the 5\% criterion, therefore, we averaged the ESS across the variance and quantile estimators only.}

\begin{figure}[htb]
	\centering
	\includegraphics[width=.45\textwidth]{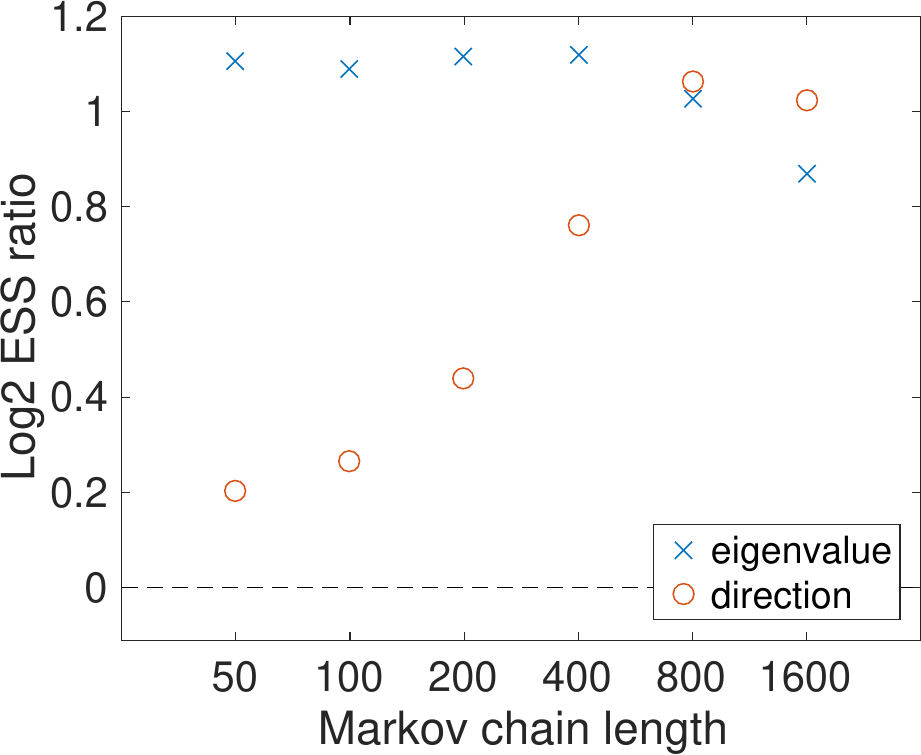} 
	\caption{Performance comparison between HMC with and without recycling in estimating the direction and magnitude of the leading principal component for the covariance matrix in the Gaussian example.}
	\label{fig:Gaussian_cov_est}
\end{figure}

The ratios of ESS in estimating the angle as well as the eigenvalues are shown in Fig.~\ref{fig:Gaussian_cov_est}. We plot the ratios against the lengths of Markov chains. The direction of the principal component cannot be well estimated by shorter chains of lengths $\sim 200$ even with recycling, but recycling conveys a substantial advantage as the chains are run longer. 
While here the ESS ratios vary substantially as a function of chain lengths, this dependence seems to be unique to this particular summary statistics --- we show in Appendix~\ref{app:ess_ratio_as_function_of_chain_length} that, for the other statistics, the magnitudes of ESS gains from recycling are independent of the chain lengths.

\begin{figure}[htb]
	\centering
	\begin{subfigure}[t]{0.45\textwidth}
		\includegraphics[width=\textwidth]{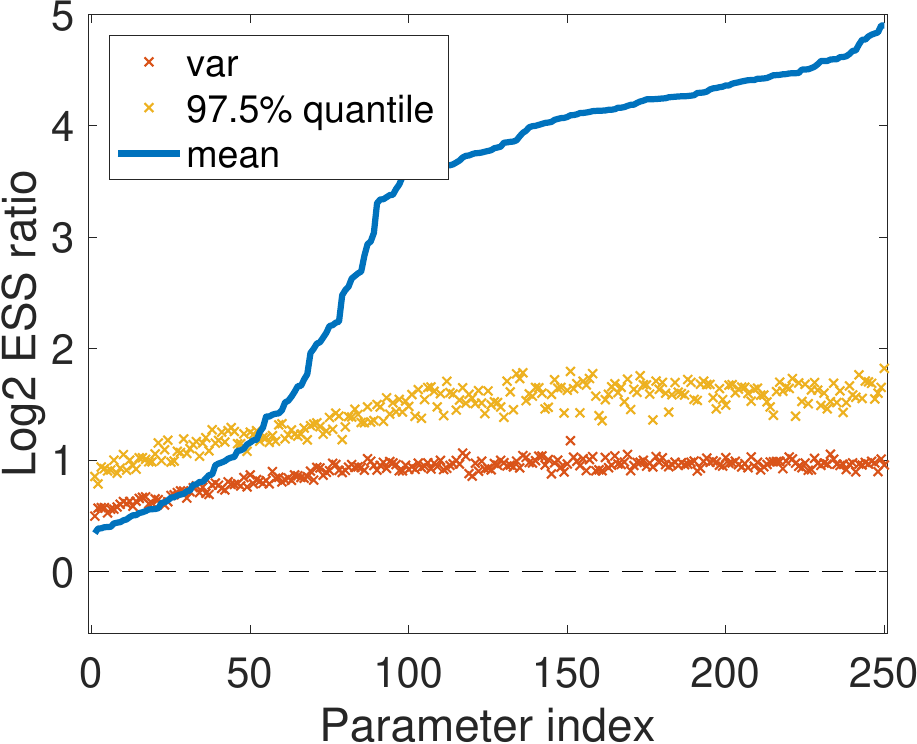} 
	\end{subfigure}
	~
	\begin{subfigure}[t]{0.45\textwidth}
		\includegraphics[width=\textwidth]{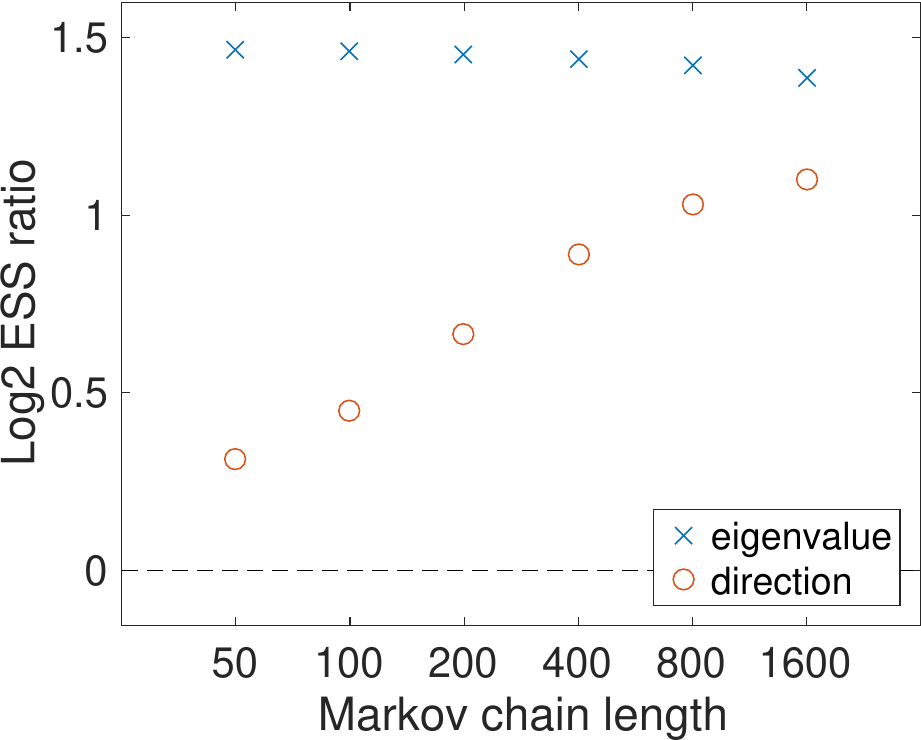} 
	\end{subfigure}
	\caption{Performance comparison between NUTS with and without recycling for the Gaussian example.}
	\label{fig:Gaussian_NUTS_comparison}
\end{figure}

\newpage
Similar performance comparisons of NUTS with and without recycling are provided in Fig.~\ref{fig:Gaussian_NUTS_comparison}. The average trajectory length was $2^9 = 512$, out of which $2^4 - 1 = 15$ samples were recycled. 

\begin{figure}[htb]
	\centering
	\includegraphics[width=.45\textwidth]{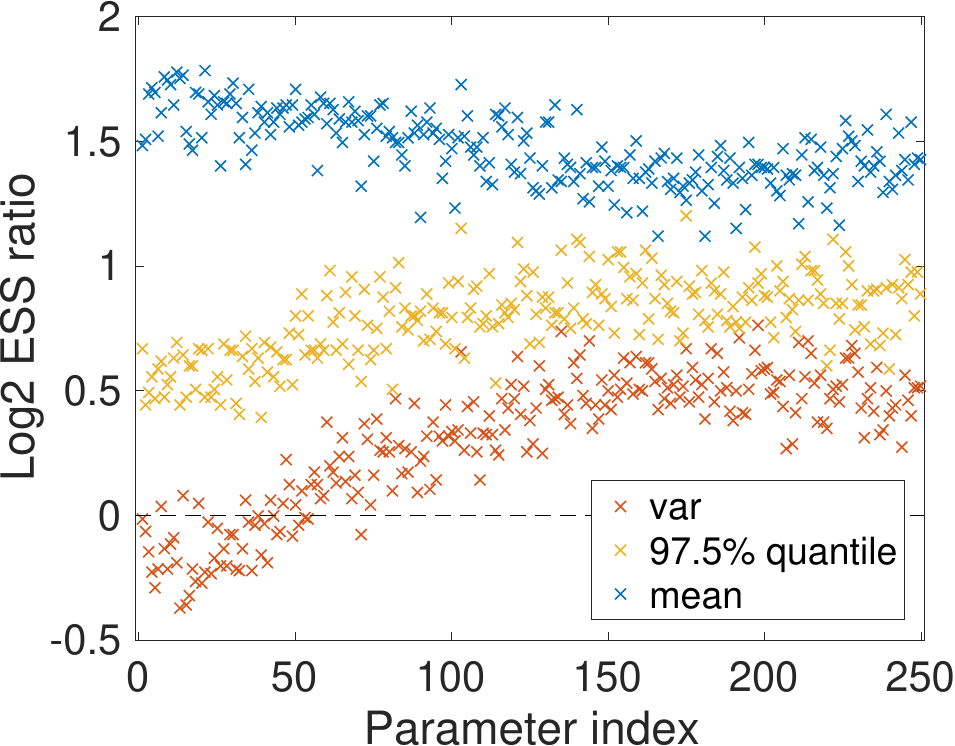} 
	\caption{$\log_2$ ratios of average ESS based on $10^4$ gradient evaluations when the mass matrix is tuned with and without recycling for $N_{\rm adap} = 400$ in the Gaussian example.}
	\label{fig:Gaussian_tuning_comparison}
\end{figure}

Lastly, the results of our covariance/mass matrix tuning experiment is summarized in Figure~\ref{fig:Gaussian_tuning_comparison}.
% shows the ratio of average ESS with and without recycling during the covariance estimation phase for $N_{\rm adap} = 400$. 
Again, in this experiment recycling is only carried out during the tuning phase and the difference in ESS comes purely from difference in the accuracy of the covariance estimators. The benefit of recycling diminishes as $N_{\rm adap}$ increases as the covariance matrix can be adequately approximated without recycling and we found no advantage of recycling when $N_{\rm adap} \geq 800$.

\FloatBarrier

\subsection{Hierarchical Bayesian logistic regression}
\label{sec:HBLR_test_case}
The second test case is a hierarchical Bayesian logistic regression model applied to the German credit data set available from the University of California Irvine Machine Learning Repository. Including two-way interaction terms and an intercept, there are 301 predictors and the regression coefficients $\bm{\beta}$ are given a $\Normal{\sigma^2 \bm{I}}$ prior. A hyper-prior is placed on $\sigma^2$, which makes the posterior inference more challenging through the strong dependence between $\sigma$ and $\bm{\beta}$. We made one modification to the corresponding example in \cite{hoffman14} by defining our parameters to be $(\log(\sigma), \bm{\beta})$ instead of $(\sigma^2, \bm{\beta})$ since such a transformation of constrained variables has become standard \citep{stan15}. A default flat prior was placed on $\sigma$.

%The 800 independent chains were run for 3200 iterations starting from stationarity. In computing the ESSs, the statistics from an independent chain of $10^7$ NUTS iterations after $10^3$ burn-in samples were used as the ground truth. 	

\begin{figure}[htb]
	\centering
	\begin{subfigure}{0.45\textwidth}
		\includegraphics[width=\textwidth]{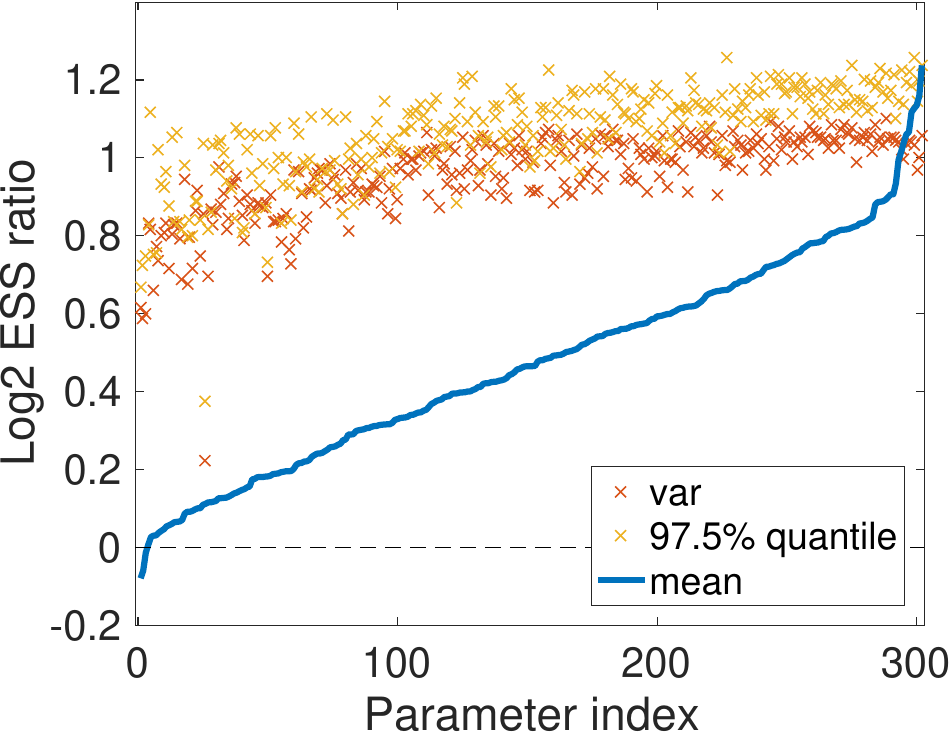} 
	\end{subfigure}
	~
	\begin{subfigure}{0.45\textwidth}
		\includegraphics[width=.96\textwidth]{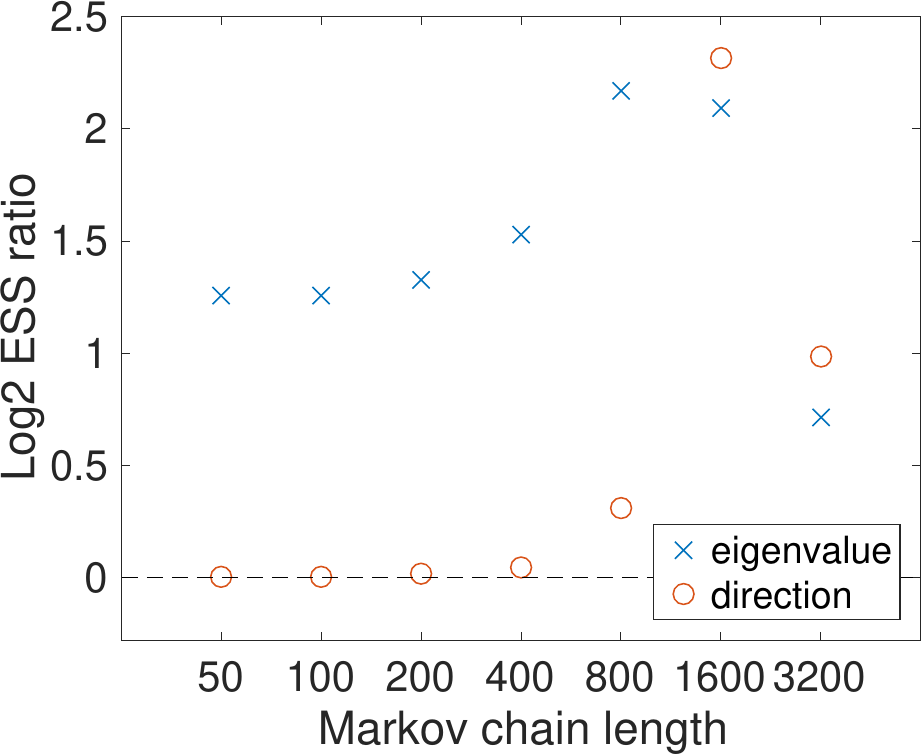} 
	\end{subfigure}
	\caption{Performance comparison between HMC with and without recycling for the hierarchical logistic model.}
	\label{fig:HBLR_comparison}
\end{figure}

\begin{figure}[htb]
	\centering
	\begin{subfigure}[t]{0.45\textwidth}
		\includegraphics[width=\textwidth]{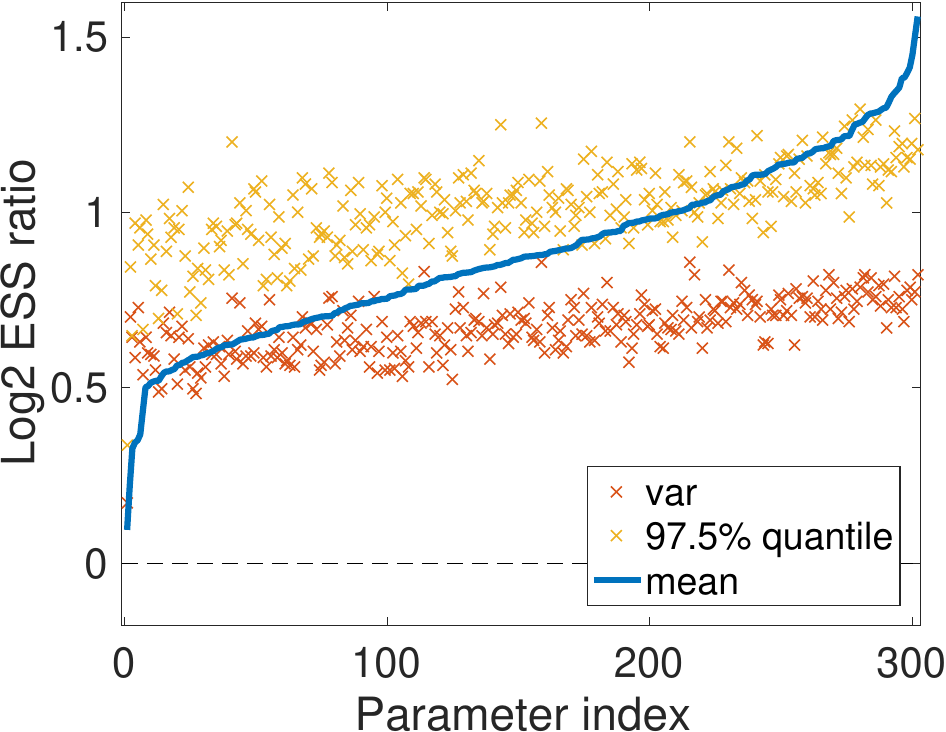} 
	\end{subfigure}
	~
	\begin{subfigure}[t]{0.45\textwidth}
		\includegraphics[width=.96\textwidth]{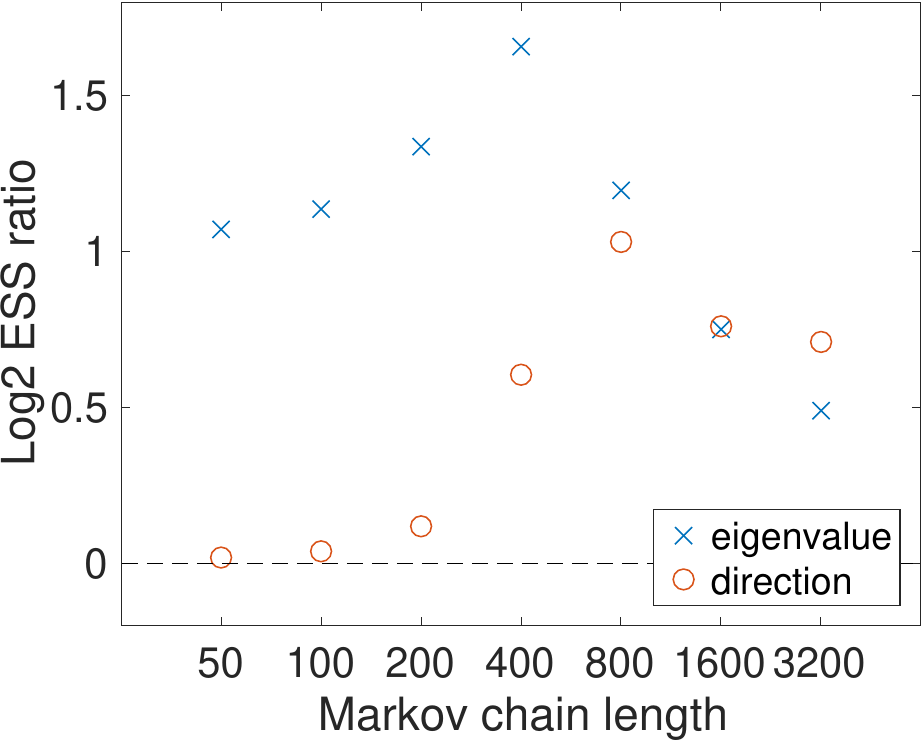} 
	\end{subfigure}
	\caption{Performance comparison between NUTS with and without recycling for the hierarchical logistic model.}
	\label{fig:HBLR_NUTS_comparison}
\end{figure}

\begin{figure}[htb]
	\centering
	\includegraphics[width=.45\textwidth]{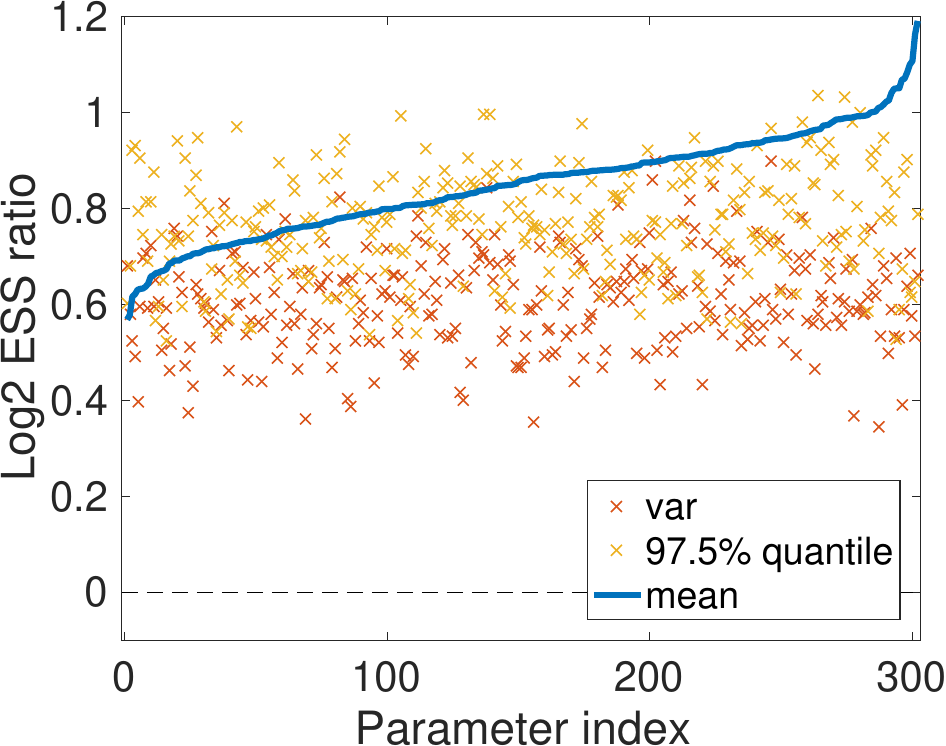} 
	\caption{Comparison of average ESS based on $10^4$ gradient evaluations between NUTS with a mass matrix tuned with and without recycling for $N_{\rm adap} = 500$ in the hierarchical logistic model.}
	\label{fig:HBLR_tuning_comparison}
\end{figure}

Performance comparisons as in the previous example are shown in Fig.~\ref{fig:HBLR_comparison}, \ref{fig:HBLR_NUTS_comparison}, and \ref{fig:HBLR_tuning_comparison}.
For some parameters, recycling seems to produce little gains in terms of mean estimation but provides clear benefits in terms of variance and quantile estimation. 
%which is critical for quantifying uncertainty of the probabilistic model. 
In the mass matrix tuning experiment shown in Figure~\ref{fig:HBLR_tuning_comparison}, we tried $N_{\rm adap} = 500, 1000, 2000$ and observed substantial improvement in the average ESS from recycling for $N_{\rm adap} \leq 1000$. 

For the path lengths for HMC, we first found the value $\tau$ to maximize the normalized expected square jumping distance $\tau^{-1/2} \mathbb{E}\| \btheta^{(i+1)}(\tau) - \btheta^{(i)}(\tau) \|$ as in \cite{wang13}, then jittered each path length $\tau^{(i)}$ in the range $[0.9 \, \tau, 1.1 \, \tau]$. The average trajectory length of HMC was 9 and all the intermediate states were recycled. The average trajectory length of NUTS was $2^4 = 16$, out of which $7$ were recycled.
% Incidentally, this optimization of path lengths can be carried out efficiently by the recycling algorithm as in Theorem~\ref{thm:vector_ergodic_theorem}, another advantage of recycled HMC. 

\FloatBarrier
\subsection{Stochastic volatility model}
\label{sec:SV_test_case}
The third test case is a stochastic volatility (SV) model fit to a time series $\bm{y}$ taken from the closing values of S{\&}P 500 index for 3000 days ending on Dec 31st, 2015. The model is specified as follows:
\begin{align*}
\log \left( \frac{y_i}{y_{i-1}} \right) 
&\sim \Normal[0]{s_i^2}, \ 
100 \log \left( \frac{s_i}{s_{i-1}} \right) 
\sim \Normal[0]{\tau^{-1}}
\end{align*}
with priors $s_0 \sim \text{Exp}(\text{mean} = 1/10)$ and $\tau \sim \text{Gamma}\left(1/2, 1/2 \right)$. The observed value on Jan 2nd, 2008 was removed from the original data as this simple SV model could not fit this observation well. 
% The model is identical to the one in \cite{hoffman14} except for minor changes to simplify the analytical formula of posterior density. 
After integrating out $\tau$ to accelerate mixing, we are left with a 3000 dimensional parameter space for $\log s$. 

\begin{figure}[htb]
	\centering
	\begin{subfigure}{0.45\textwidth}
		\includegraphics[width=\textwidth]{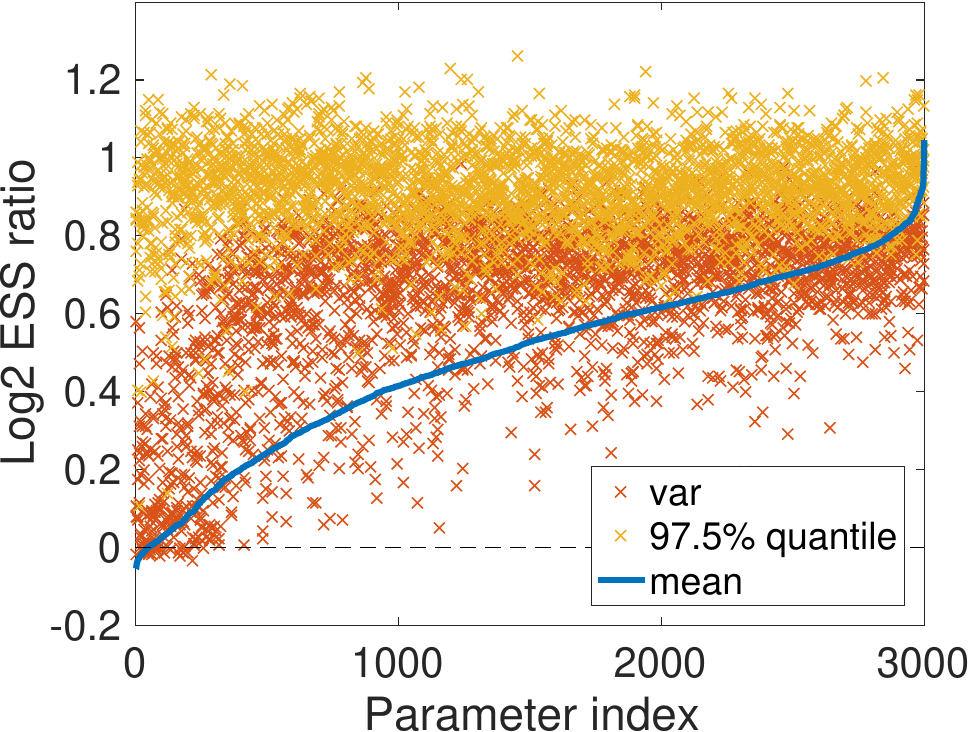} 
	\end{subfigure}
	~
	\begin{subfigure}{0.45\textwidth}
		\includegraphics[width=.96\textwidth]{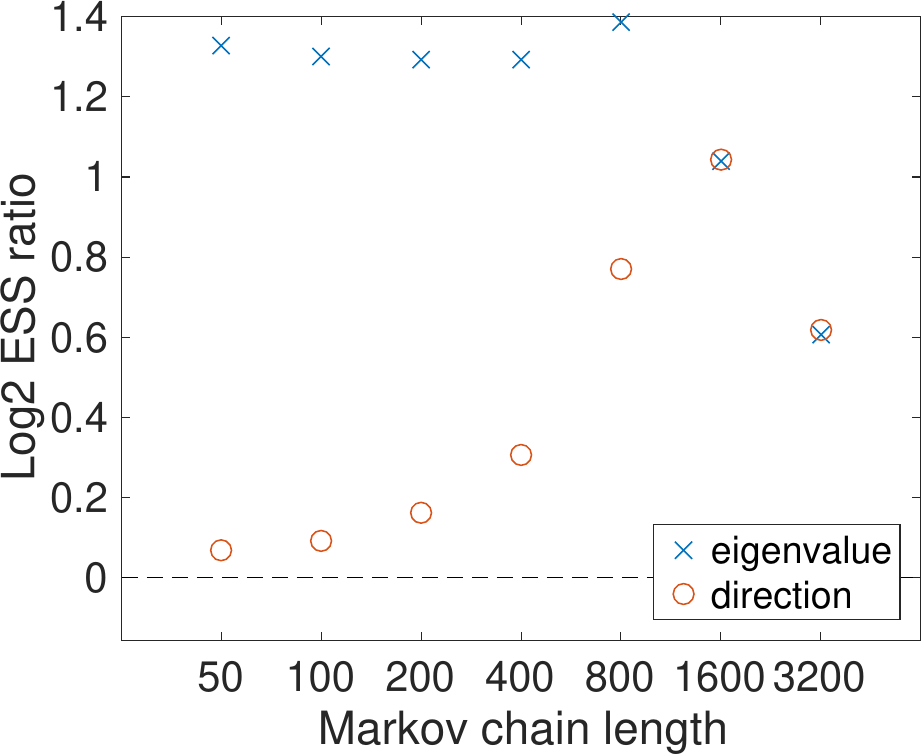} 
	\end{subfigure}
	\caption{Performance comparison between HMC with and without recycling for the SV model.}
	\label{fig:SV_comparison}
\end{figure}

\begin{figure}[htb]
	\centering
	\begin{subfigure}{0.45\textwidth}
		\includegraphics[width=\textwidth]{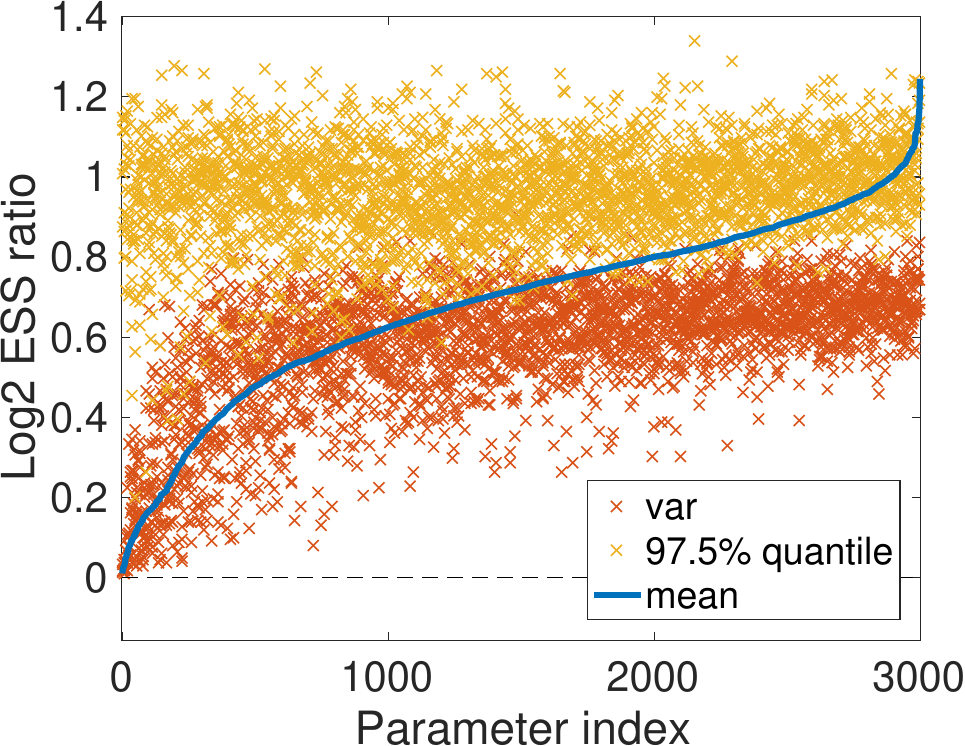} 
	\end{subfigure}
	~
	\begin{subfigure}{0.45\textwidth}
		\includegraphics[width=.96\textwidth]{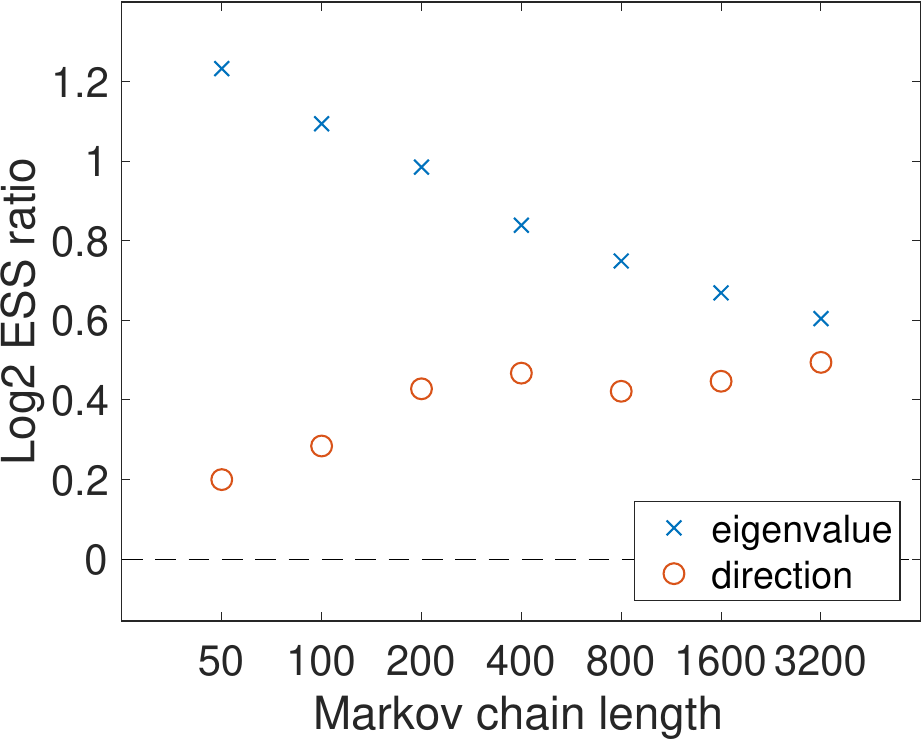} 
	\end{subfigure}
	\caption{Performance comparison between NUTS with and without recycling for the SV model.}
	\label{fig:SV_NUTS_comparison}
\end{figure}

Performance comparisons as in the previous examples are shown in Fig.~\ref{fig:SV_comparison} and \ref{fig:SV_NUTS_comparison}.
% The 400 independent chains were run for 3200 iterations starting from stationarity. In computing the ESSs, the statistics from an independent NUTS chain of length $2.5 \times 10^6$ after $10^3$ burn-in were used as the ground truth. 
The path length for HMC was chosen as in Section~\ref{sec:HBLR_test_case}. 
% The mass matrix tuning experiment was not carried out for this example as tuning a mass matrix for a $3000$ dimensional space is impractical. 
On average, 44 samples out of 90 per iteration were recycled for HMC and 7 out of $2^7 = 128$ were recycled for NUTS.

\subsection{Log-Gaussian Cox point-process (Riemann manifold HMC)}
\label{sec:logGaussCox_test_case}
The last test case is a log-Gaussian Cox point-process model from \cite{girolami11}, where they apply \textit{Riemann manifold HMC} (RMHMC) to sample from the latent process $\bm{x} \in \mathbb{R}^{4096}$ defined on a $64 \times 64$ grid. The observation $y_{ij}$ for $i , j = 1, \ldots, 64$ is assumed to follow 
\begin{equation}
y_{ij} \given x_{ij} 
	\sim \textrm{Poisson}(\exp(x_{ij}) / 64^2).
\end{equation}
The latent process $\bm{x} \given \mu, \sigma, \ell$ is given a Gaussian process prior with mean $\mu \bm{1}$ and covariance 
\begin{equation}
\textrm{cov}(x_{ij}, \, x_{i'j'})
	= \sigma^2 \exp\!\left(
		\frac{
			- \sqrt{(i - i')^2 + (j - j')^2}
		}{
			64 \ell
		}
	\right).
\end{equation}
Following \cite{girolami11}, we fix the hyper-parameters at $\ell = 1 / 33$,  $\sigma^2 = 1.91$, and $\mu = \log(126) - \sigma^2 / 2$ and simulate the data from the generative model.

Due to the greater computational cost of simulation in this example, we only run the 100 independent chains for 1,600 iterations. The ground truth statistics are calculated from a NUTS chain of $10^6$ iterations.\footnote{NUTS is actually a meta-algorithm that provides a useful trajectory termination criterion for any MCMC algorithm based on reversible dynamics. In particular, NUTS and our recycled version apply straightforwardly to most of the HMC variants, including RMHMC.}
The path length of RMHMC is uniformly sampled from the range 1 to 30 as done in \cite{girolami11}. 

\begin{figure}[htb]
	\centering
	\begin{subfigure}{0.45\textwidth}
		\includegraphics[width=\textwidth]{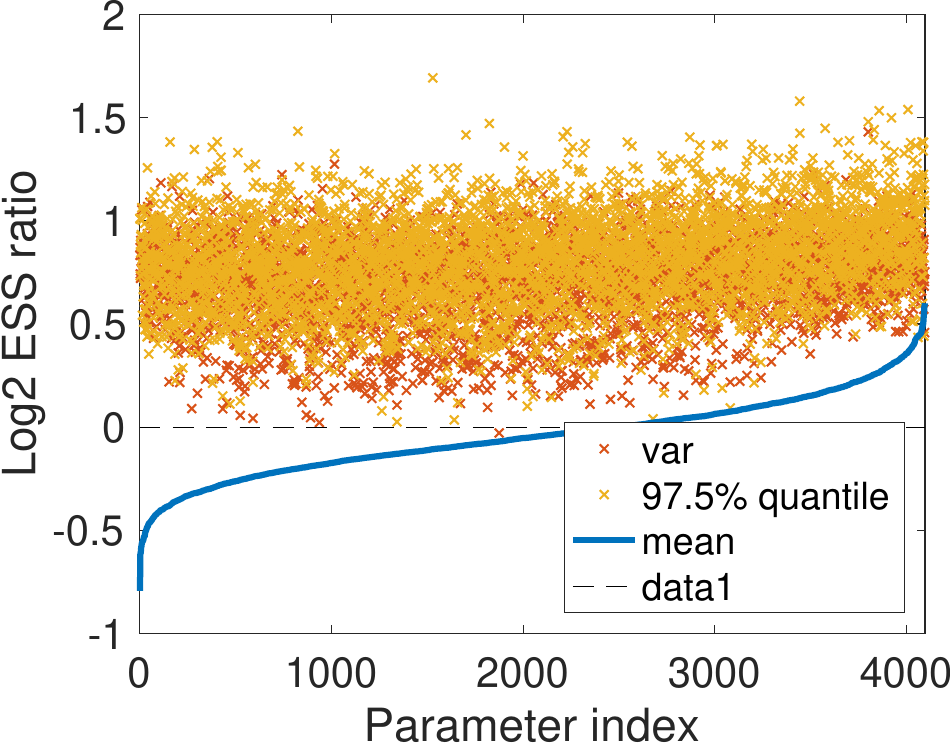} 
	\end{subfigure}
	~
	\begin{subfigure}{0.45\textwidth}
		\includegraphics[width=.96\textwidth]{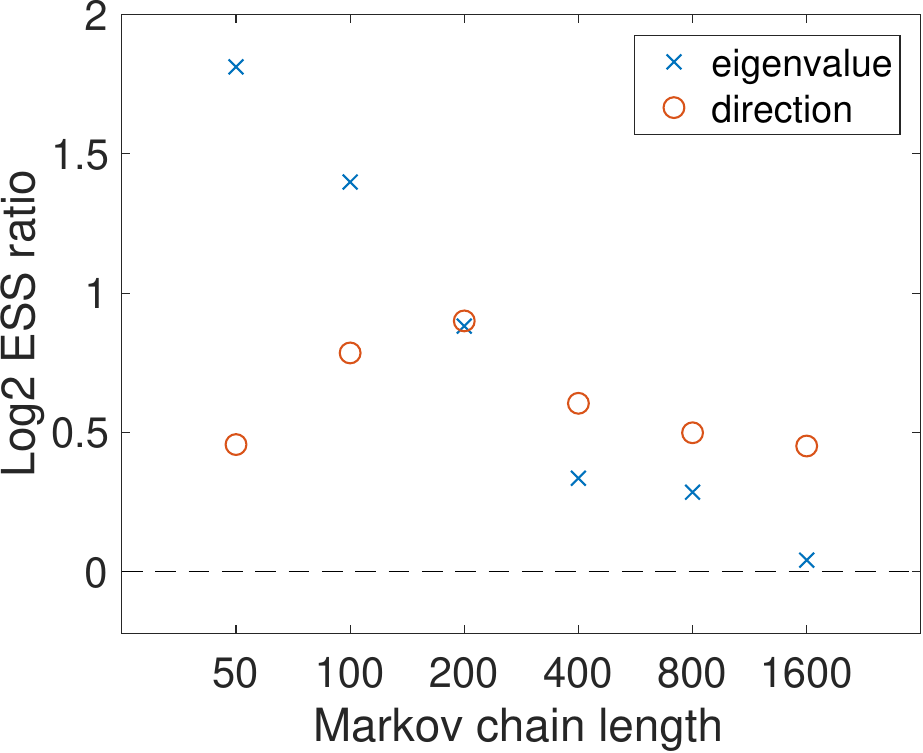} 
	\end{subfigure}
	\caption{Performance comparison between RMHMC with and without recycling for the log-Gaussian Cox model.}
	\label{fig:logGaussCox_comparison}
\end{figure}

\begin{figure}[htb]
	\centering
	\begin{subfigure}{0.45\textwidth}
		\includegraphics[width=\textwidth]{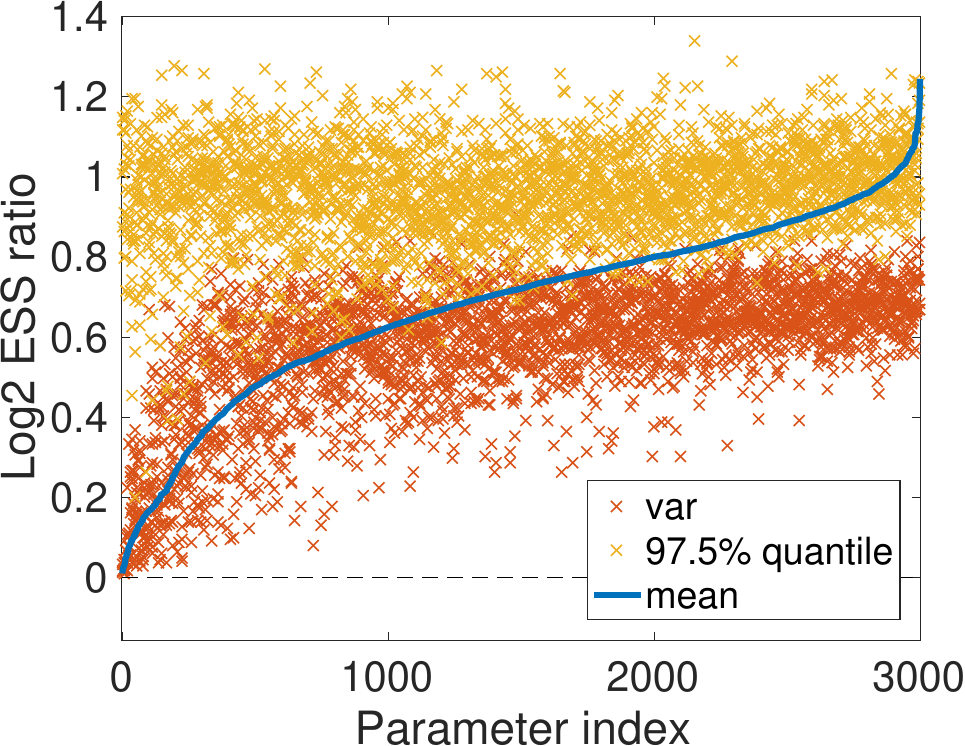} 
	\end{subfigure}
	~
	\begin{subfigure}{0.45\textwidth}
		\includegraphics[width=.96\textwidth]{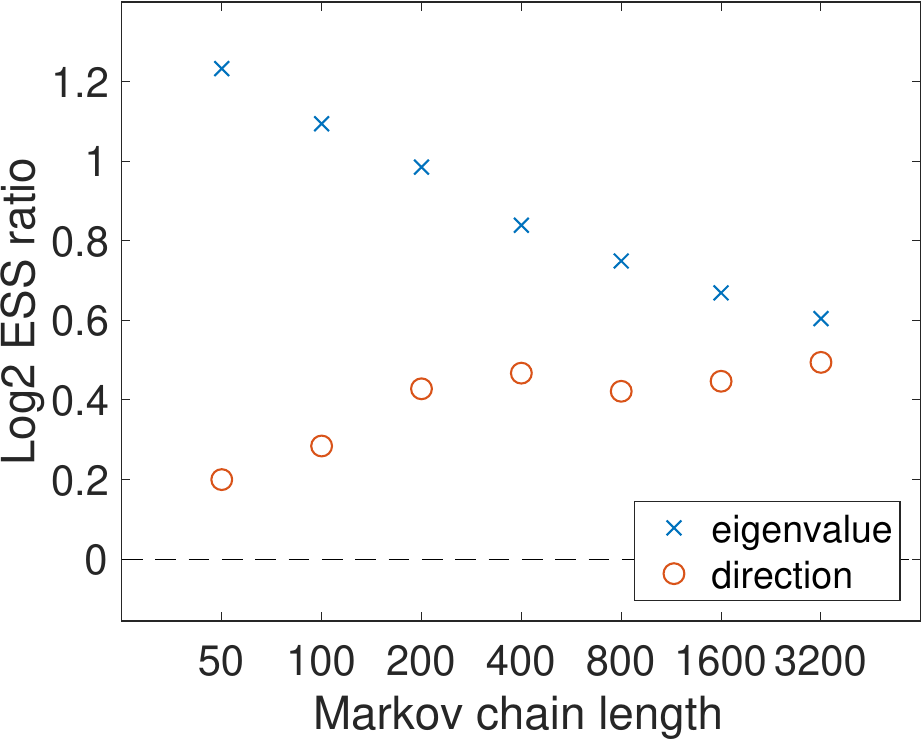} 
	\end{subfigure}
	\caption{Performance comparison between NUTS with and without recycling for the log-Gaussian Cox model.}
	\label{fig:logGaussCox_NUTS_comparison}
\end{figure}

Performance comparisons as in the previous examples are shown in Fig.~\ref{fig:logGaussCox_comparison} and \ref{fig:logGaussCox_NUTS_comparison}. 
Recycled RMHMC yields worse ESS for the majority of the mean estimators in this example, but this is an artifact of the negative auto-correlations in non-recycled RMHMC causing \textit{super-efficiency} --- MCMC samples yielding smaller Monte Carlo errors than the same number of independent samples.
% \citep{thawornwattana2018designing-proposal}. 
HMC, and hence RMHMC, is known to occasionally exhibit such behavior \citep{kennedy1991hmc-acceptprob-and-autocorr, neal10}.
We can see, however, that recycling uniformly and significantly improves ESS for the variance and quantile estimators. Under NUTS, such super-efficiency artifacts are not observed and recycling improves ESS for all the estimators.
%The mass matrix tuning experiment again is not carried out for this example due to the high-conditionality of the parameter space.
On average, 15 samples were recycled for RMHMC and 7 out of $2^4 = 16$ for NUTS.

\FloatBarrier
\subsection{Number of recycled samples and statistical efficiency}
\label{sec:how_many_to_recycle}
As mentioned earlier, in the simulation results above we recycle enough of the intermediate states to achieve near-optimal efficiency gains. Here we take a closer look at how the efficiency gain from recycling depends on the number of recycled samples. The results in particular provide a practical guidance on how one might trade off statistical efficiency for memory efficiency when the available memory becomes limited. 

\begin{figure}[htb]
	\centering
	\begin{subfigure}{0.45\textwidth}
		\includegraphics[width=\textwidth]{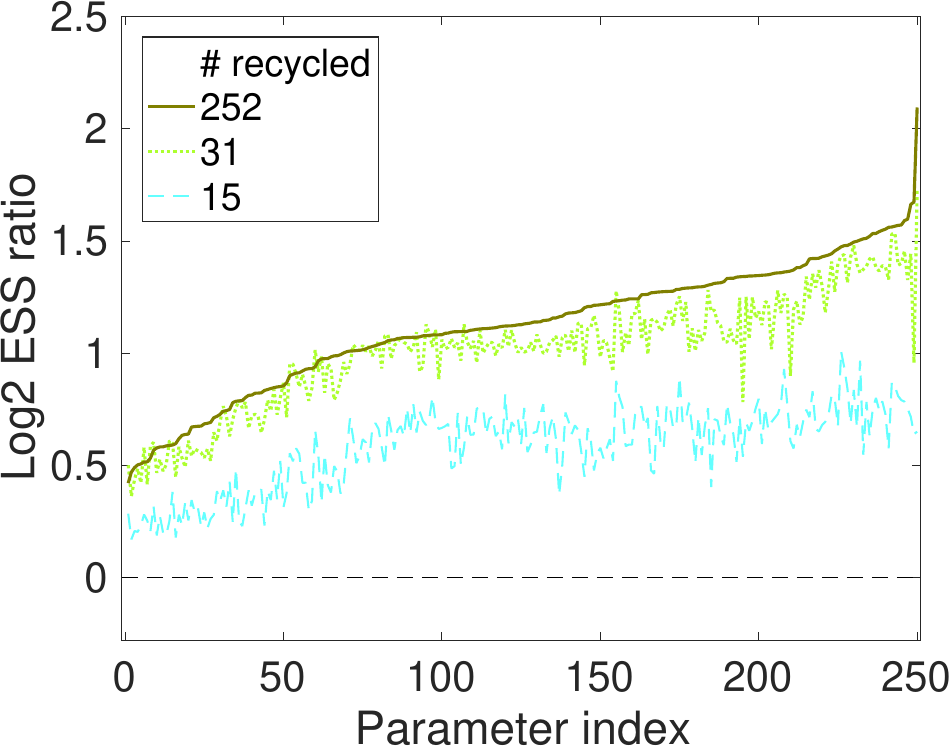} 
	\end{subfigure}
	~
	\begin{subfigure}{0.45\textwidth}
		\includegraphics[width=\textwidth]{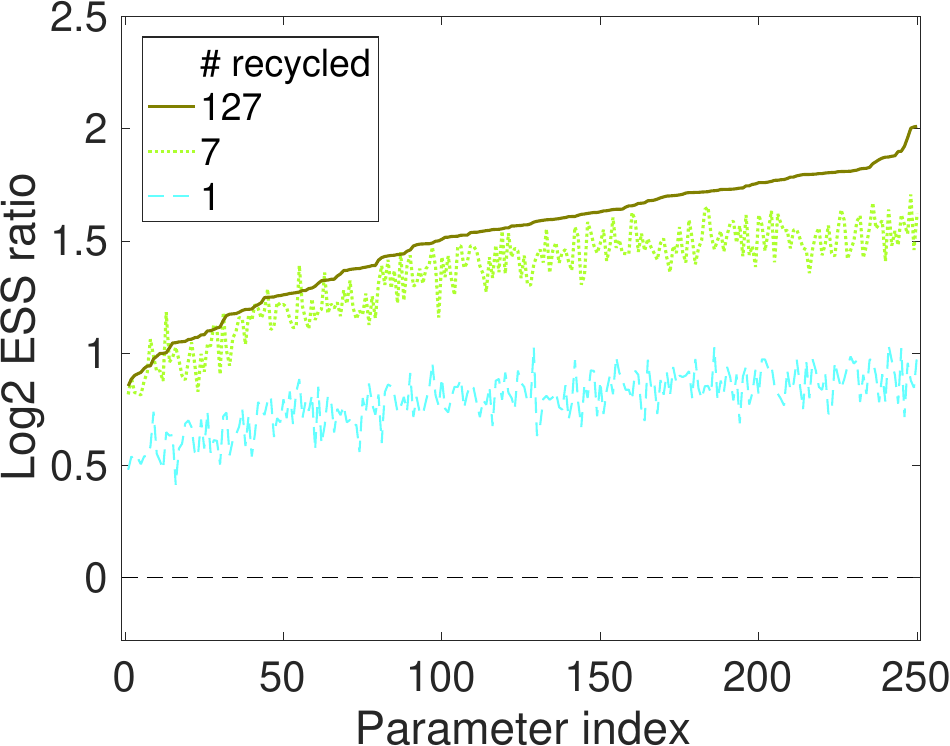} 
	\end{subfigure}
	\caption{Multivariate Gaussian example: improvement in ESS for 97.5\% quantile estimation with different number of recycled samples.}
	\label{fig:Gaussian_nrecycle_experiment}
\end{figure}

\begin{figure}[htb]
	\centering
	\begin{subfigure}{0.45\textwidth}
		\includegraphics[width=\textwidth]{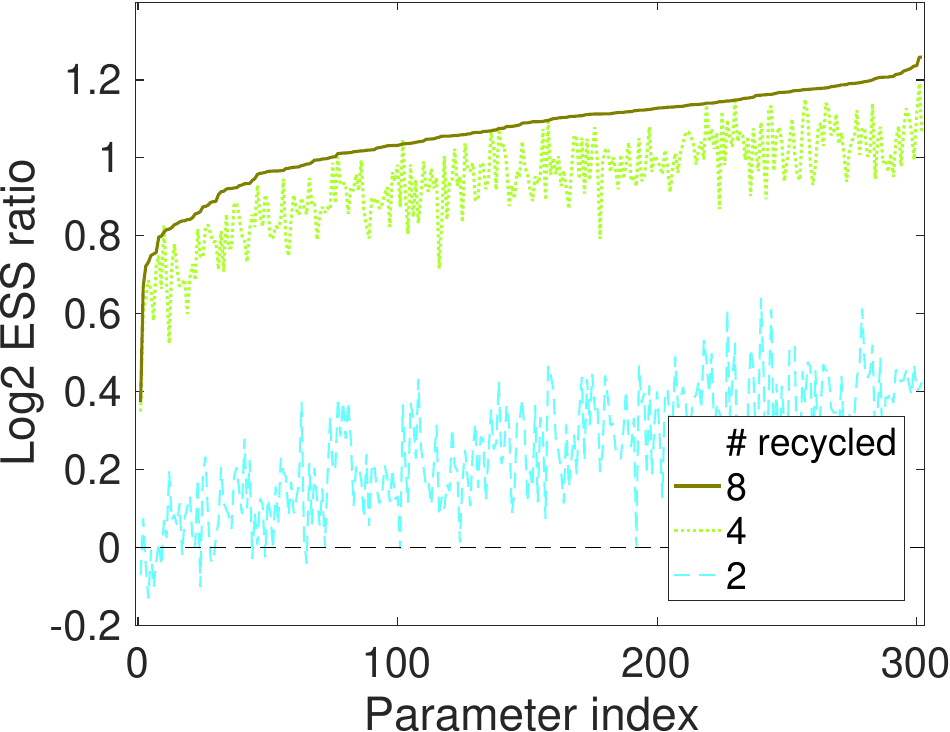} 
	\end{subfigure}
	~
	\begin{subfigure}{0.45\textwidth}
		\includegraphics[width=\textwidth]{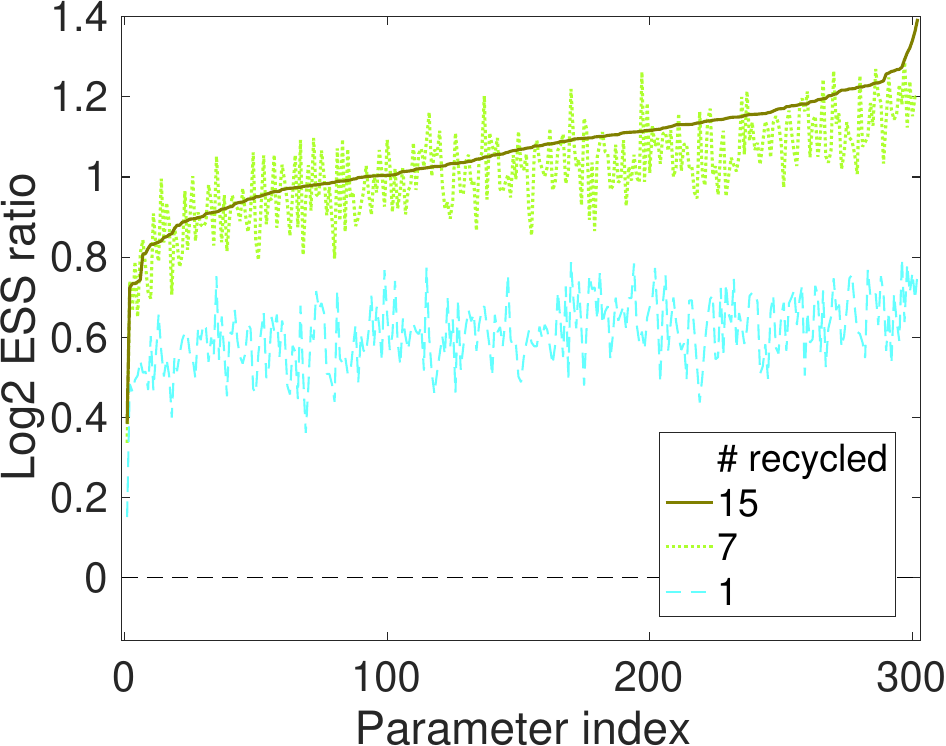} 
	\end{subfigure}
	\caption{Hierarchical logistic example: improvement in ESS for 97.5\% quantile estimation with different number of recycled samples.}
	\label{fig:HBLR_nrecycle_experiment}
\end{figure}

\begin{figure}[htb]
	\centering
	\begin{subfigure}{0.45\textwidth}
		\includegraphics[width=\textwidth]{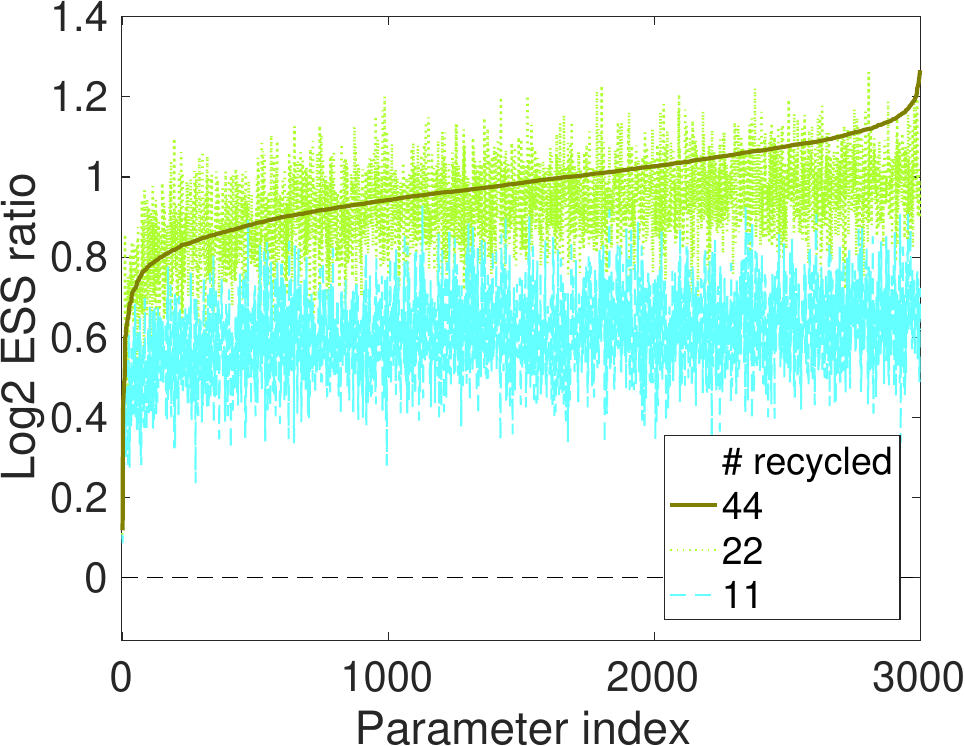} 
	\end{subfigure}
	~
	\begin{subfigure}{0.45\textwidth}
		\includegraphics[width=\textwidth]{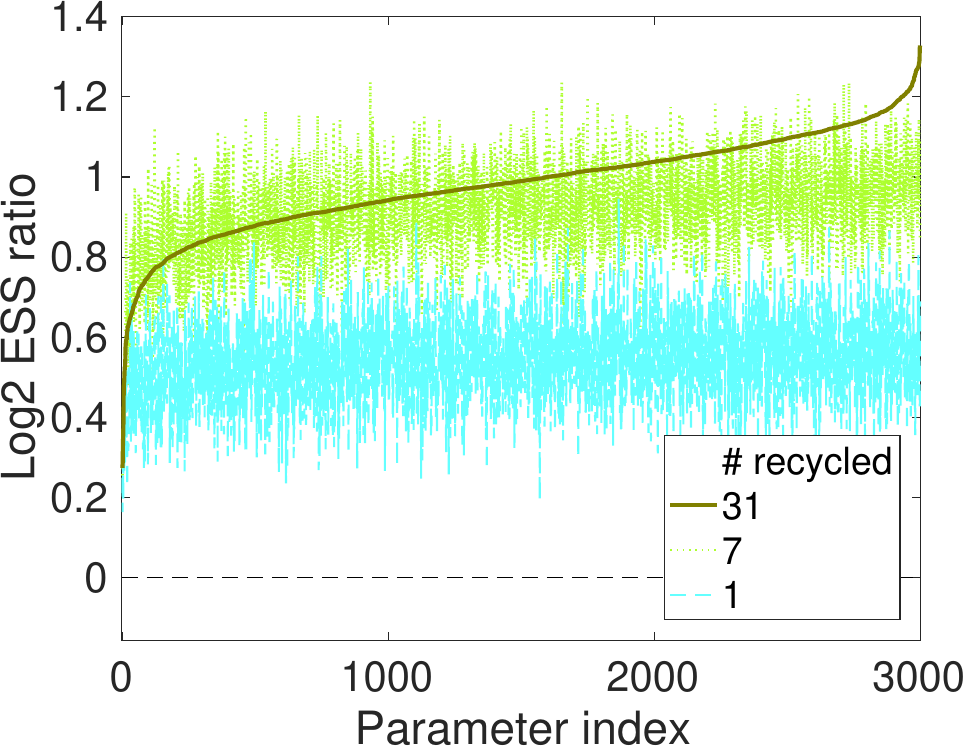} 
	\end{subfigure}
	\caption{Stochastic volatility example: improvement in ESS for 97.5\% quantile estimation with different number of recycled samples.}
	\label{fig:SV_nrecycle_experiment}
\end{figure}

\begin{figure}[htb]
	\centering
	\begin{subfigure}{0.45\textwidth}
		\includegraphics[width=\textwidth]{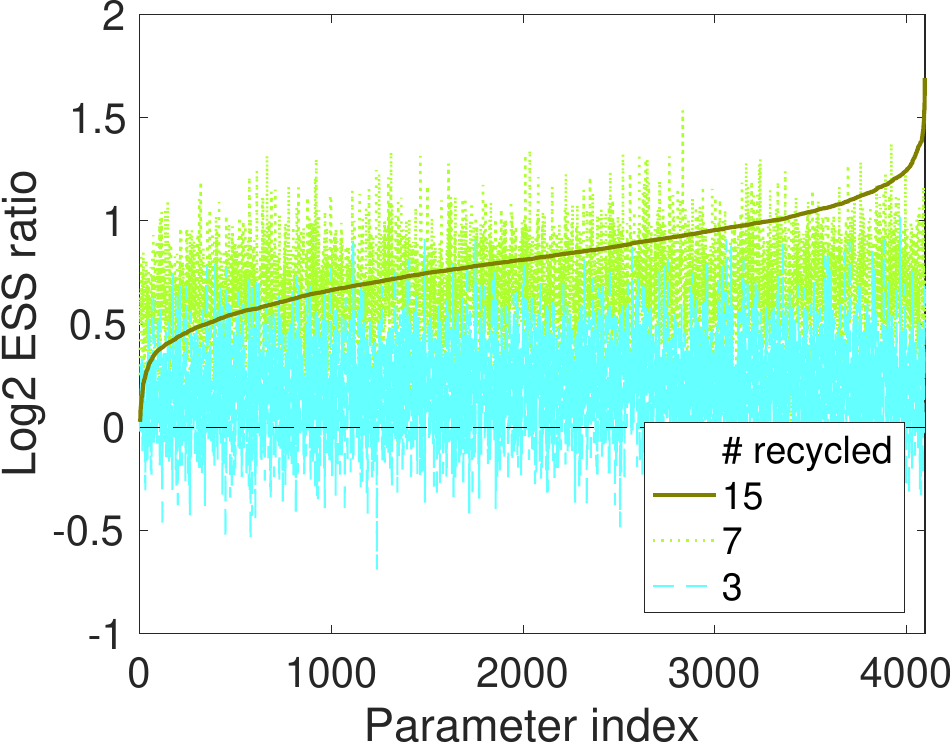} 
	\end{subfigure}
	~
	\begin{subfigure}{0.45\textwidth}
		\includegraphics[width=\textwidth]{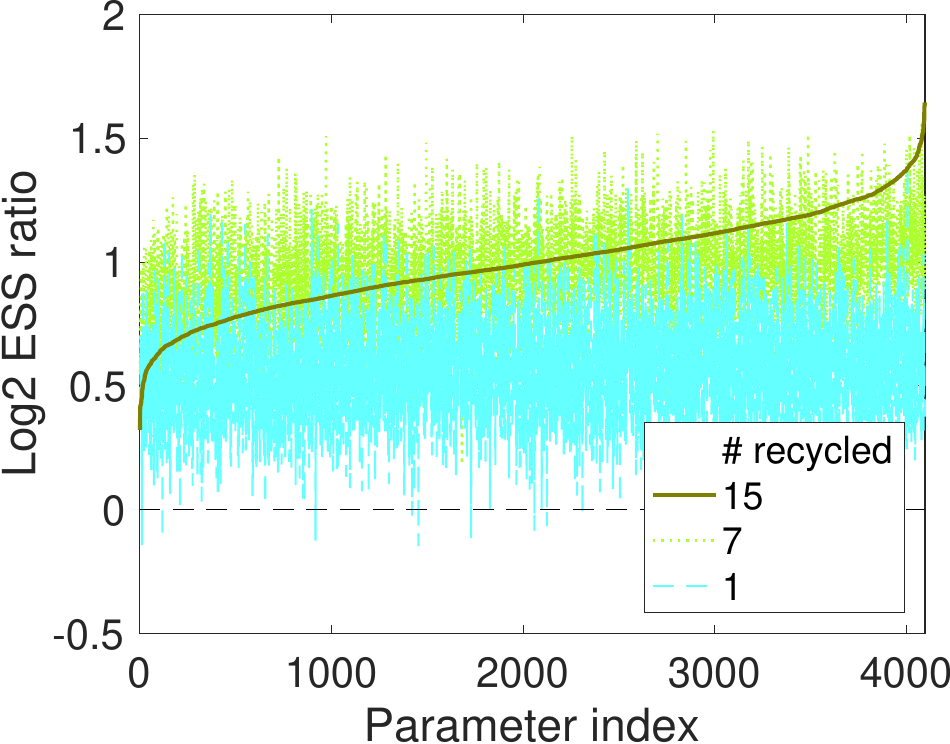} 
	\end{subfigure}
	\caption{Log-Gaussian Cox example: improvement in ESS for 97.5\% quantile estimation with different number of recycled samples.}
	\label{fig:logGaussCox_nrecycle_experiment}
\end{figure}

For our experiments here, we focus on the problem of estimating a quantile; the dependence of mean and variance estimators on the number of recycled samples was found to be similar. The number of samples per iteration was repeatedly reduced by a factor of 2 until the benefit of recycling became almost negligible. The results are summarized in the $\log_2$ ESS ratio plots as presented earlier; Figure~\ref{fig:Gaussian_nrecycle_experiment} for the multi-variate Gaussian example, Figure~\ref{fig:HBLR_nrecycle_experiment} for the hierarchical Bayesian logistic regression example, Figure~\ref{fig:SV_nrecycle_experiment} for the stochastic volatility example, and Figure~\ref{fig:logGaussCox_nrecycle_experiment} for the log-Gaussian Cox example.
% The parameter indices are sorted in the increasing order of the ESS ratio at the largest number of recycled samples (the dark solid line). 
The green dotted line corresponds to the number of recycled samples at which the efficiency decrease relative to the optimal one becomes visually noticeable. The cyan dashed line corresponds to the number of recycled samples below which the benefit from recycling becomes negligible. 

The performance of recycled NUTS is particularly remarkable, not only offering the near-optimal efficiency gain well-below the maximal recycling size but also demonstrating over 40\% ($\approx 2^{0.5}$) efficiency gain with just one recycled sample. For recycled HMC, the efficiency gains remain substantial well-below the maximal recycling size but start to diminish much earlier than NUTS. Two design features of NUTS likely explain this phenomenon. First, NUTS simulates a trajectory in both the forward and backward direction, which means that some of the intermediate states lie in the direction opposite to the final proposal state relative to the starting point of a trajectory.  Secondly, while HMC simulates a trajectory to construct one high-quality proposal state, NUTS generates a collection of states --- any of which likely constitutes a good proposal state --- and selects one state from the collection as a final proposal. 
Compared to those of HMC, therefore, the recyclable states of NUTS probably have smaller correlations with the final proposal state. 
% Even if the efficiency gain is comparable between HMC and NUTS when recycling all the intermediate states, it seems that the smaller pair-wise correlations of recyclable states with the final state provides NUTS with a greater benefit when recycling a small subset.

Our experiments here suggest that recycled NUTS may be a particularly practical alternative to the standard implementation of HMC-type algorithms; it not only eliminates the need to tune the path length but also provides a significant boost in efficiency with a rather small increase in memory requirement. 

%These two features of NUTS suggest that, compared to that of HMC, a randomly chosen recyclable state of NUTS have a relative smaller correlation with the final state.

%Our findings here suggest that any amount of recycling would improve NUTS and that a near-optimal efficiency can be often achieved with less than 10 recycled samples. 

\newpage
\section{Discussion}

We have proposed a simple and general algorithm for improving the efficiency of HMC and variants with essentially no extra computational overhead. 
%The trade-off between statistical and memory efficiency was also studied as it is an important scalability issue. 
Our simulations demonstrate the substantial gains in computational efficiency without excessive memory use. In practice, conceptual complexity, ease of implementation, and memory efficiency are just as important considerations as statistical efficiency.
These considerations can explain why related ideas to improve the efficiency of HMC variants have not gained traction. Our algorithm provides a more practical and user-friendly alternative that applies straightforwardly to a wide range of multi-proposal schemes.

\newpage
\FloatBarrier
%% ** The bibliograhy **
\bibliographystyle{ba}
\bibliography{recycled_hmc}{} % place <bib-data-file> 

\newpage
\appendix

\section{Recycling varying numbers of states}
\label{app:recycling_variable_number_of_states}

Theorem~\ref{thm:variable_size_vector_ergodic_theorem} below extends Theorem~\ref{thm:vector_ergodic_theorem} to allow recycling the intermediate states of a multi-proposal scheme in which the number of generated states varies randomly from one iteration to another. Theorem~\ref{thm:general_ReHMC} is a special case of Theorem~\ref{thm:variable_size_vector_ergodic_theorem} when the transition kernel $P_k(\cdot \given \cdot)$ is chosen as one iteration of HMC with $k$ leapfrog steps for $k \geq 1$ and $P_0(\cdot \given \cdot)$ as that with $L \sim \pi_\Lambda(\cdot)$ steps.

\begin{theorem} 
	\label{thm:variable_size_vector_ergodic_theorem}
	Let $L \sim \pi_\Lambda(\cdot)$ denote a random variable taking values on $\{1, \ldots, K\}$. Let $P_k(\cdot \given \cdot)$ for $k = 0, 1, \ldots, K$ be transition kernels with a common stationary measure $\pi(\cdot)$ on a space $\Omega$ and suppose $P_0(\cdot \given \cdot)$ is uniquely ergodic.
	Denote by $(\chainState, \ell) = (\chainState_0, \ldots, \chainState_\ell, \ell)$ an element of a space $\cup_{\ell = 1}^K \Omega^{\ell + 1} \times \{\ell\}$.
	Consider a Markov chain $\{(\chainState^{(i)}, L^{(i)})\}_{i \geq 1}$ whose transition probability $(\chainState, L) \to (\chainState^*, L^*)$ satisfies
	\begin{equation} 
	\label{eq:assump_variable_length_first_coord_only}
	P(\chainState^*, L^* = \ell \given \chainState, L)
	= Q_{\ell}(\chainState^*_0, \ldots, \chainState^*_{\ell} \given \chainState_0) \, \pi_\Lambda(L^* = \ell),
	\end{equation}	
	where each $Q_\ell(\cdot \given \chainState_0)$ has the marginal densities
	\begin{equation}
	\label{eq:assump_variable_length_marginal}
	\int Q_\ell(\chainState^*_0, \ldots, \chainState^*_{\ell}  \given \chainState_0) \, {\rm d} \chainState^*_{-k}
	= P_k(\chainState^*_k \given \chainState_0).
	\end{equation}	
%	where $\chainState^*_{-k} = (\chainState^*_0, \ldots, \chainState^*_{k-1}, \chainState^*_{k+1}, \ldots, \chainState^*_{\ell^*})$ for $k = 0, 1, \ldots, \ell$. 
	Then we have
	\begin{equation}
	\label{eq:variable_length_vector_ergodic_theorem}
	\frac{1}{\sum_{i=1}^N L^{(i)}} \sum_{i=1}^N \sum_{k =1}^{L^{(i)}} \delta_{\chainState_k^{(i)}} (\cdot) \overset{w}{\to} \pi(\cdot)
	\ \text{ as } \ N \to \infty.
	\end{equation}
	Additionally, the Markov chain $\{(\chainState^{(i)}, L^{(i)})\}_{i \geq 1}$ is geometrically (or uniformly) ergodic if $P_0(\cdot \given \cdot)$ is geometrically (or uniformly) ergodic.
\end{theorem}

\begin{proof} % [Theorem~\ref{thm:variable_size_vector_ergodic_theorem}]
	The Markov chain $(\chainState^{(1)}, L^{(1)}), (\chainState^{(2)}, L^{(2)}), \ldots$ has a stationary distribution $\pi^*(\chainState, \ell) = \pi_\ell^*(\chainState_0, \ldots, \chainState_\ell) \, \pi_\Lambda(\ell)$ where
	\vspace{-.8ex}
	\begin{equation}
	\label{eq:variable_length_vector_stationary_dist}
	\vspace{-.6ex}
	\pi_\ell^*(\chainState_0, \ldots, \chainState_\ell)
		= \int Q_\ell(\chainState_0, \ldots, \chainState_\ell \given \chainState_0) \pi(\chainState_0) \, {\rm d} \chainState_0.
	\end{equation}
	Following the same arguments as in the proof of Theorem~\ref{thm:vector_ergodic_theorem}, we can verify that \eqref{eq:variable_length_vector_stationary_dist} is in fact the unique ergodic distribution. 
	The ergodicity in particular implies $N^{-1} \sum_{i = 1}^N L^{(i)} \to \mathbb{E}_{\Lambda} \!\left[ L \right]$ almost surely, so that the empirical measure \eqref{eq:variable_length_vector_ergodic_theorem} has the same stationary distribution as
	$ N^{-1} \sum_{i=1}^N \mathbb{E}_{\Lambda}[L]^{-1} \sum_{k =1}^{L^{(i)}} \delta_{\chainState_k^{(i)}} (\cdot)$.
	To see that the preceding empirical measure converges to $\pi(\cdot)$, we only need to observe that the following identity holds for a function $f: \Omega \to \mathbb{R}$;
%	an expectation with respect to the stationary distribution $\pi^*(\cdot)$:
%	\begin{equation}
%	\mathbb{E}_{\pi^*}\!\left[
%		\frac{1}{\mathbb{E}_{\Lambda}[L]} \sum_{k =1}^{L} f(\chainState_k)
%	\right]
%		= \sum_{\ell = 1}^K \frac{\pi_\Lambda(\ell)}{\mathbb{E}_{\Lambda}[L]} \,
%			\mathbb{E}_{\pi^*_\ell} \!\left[ 
%				\sum_{k =1}^{\ell} f(\chainState_k) 
%			\right]
%		= \sum_{\ell = 1}^K \frac{\ell \, \pi_\Lambda(\ell)}{\mathbb{E}_{\Lambda}[L]} \,
%		\mathbb{E}_{\pi} \!\left[ f(\chainState_0) \right]
%		= \mathbb{E}_{\pi} \!\left[ f(\chainState_0) \right]
%	\end{equation}
	\begin{equation}
	\mathbb{E}_{\pi^*}\!\left[
		\sum_{k =1}^{L} f(\chainState_k)
	\right]
		= \sum_{\ell = 1}^K \pi_\Lambda(\ell) \,
		\mathbb{E}_{\pi^*_\ell} \!\left[ \sum_{k =1}^{\ell} f(\chainState_k) \right]
		= \sum_{\ell = 1}^K \ell \, \pi_\Lambda(\ell) \,
		\mathbb{E}_{\pi} \!\left[ f(\chainState_0) \right]
		= \mathbb{E}_{\Lambda} \!\left[ L \right] \mathbb{E}_{\pi} \!\left[ f(\chainState_0) \right]\!. 
		\qedhere
	\end{equation}
\end{proof}

\section{Simple Proof of Algorithm by Calderhead and Bernton et.\ al.\ }
\label{app:simple_proof_of_calderhead}

Here we describe how Theorem~\ref{thm:vector_ergodic_theorem} provides an alternative and simpler proof for a version of the algorithms by \cite{calderhead14} and \cite{bernton15}. The proof in particular requires no understanding of the super-detailed balance condition \citep{frenkel04, tjelmeland04}. The algorithms below are presented as ``Version 2'' of modified Calderhead's algorithms in \cite{bernton15}. As before, the map $\Fe: (\btheta_0, \momentum_0) \to (\btheta_1, \momentum_1)$ corresponds to one leap-frog step with stepsize $\epsilon$.
\begin{algo}[Calderhead and Bernton et.\ al.] 
	\label{alg:calderhead}
	Generate a Markov chain $\{ (\btheta_0^{(i)}, \momentum_0^{(i)}) \}_{i \geq 1}$ with the transition rule $(\btheta_0^{(i)}, \momentum_0^{(i)}) \to (\btheta_0^{(i+1)}, \momentum_0^{(i+1)})$ as follows:
	\begin{enumerate}
		\item Sample $L^{(i)} \sim {\rm Uniform}(\{0, 1, \ldots, K\})$. Set $\ell = K - L^{(i)}$ if $K - L^{(i)} \geq L^{(i)}$ and $\ell = - L^{(i)}$ otherwise.
		\item Set $(\btheta_0^{(i+1)}, \momentum_0^{(i+1)}) = \Fe^\ell(\btheta_0^{(i)}, \momentum_0^{(i)})$ with probability
		\begin{equation}
		\min \left\{1, 
		\frac{
			\pi\big( \Fe^\ell(\btheta_0^{(i)}, \momentum_0^{(i)}) \big)
		}{
			\pi \big( (\btheta_0^{(i)}, \momentum_0^{(i)}) \big)
		} \right\} 
		\end{equation}
		and $(\btheta_0^{(i+1)}, \momentum_0^{(i+1)}) = (\btheta_0^{(i)}, \momentum_0^{(i)})$ otherwise.
		\item Generate a new momentum: $\momentum_0^{(i+1)} \sim \Normal{\Mass}$.
	\end{enumerate}
	Additionally at each iteration, generate $\{ (\btheta_k^{(i+1)}, \momentum_k^{(i+1)}), k = 1, \ldots, K\}$ as follows:
	\begin{enumerate}
		\setcounter{enumi}{3}
		\item Define a collection of states 
		\begin{equation}
		\label{eq:calderhead_proposal_states}
		\mathcal{A}_{i+1} 
		= \mathcal{A}(\btheta_0^{(i)}, \momentum_0^{(i)}, L^{(i)})
		= \{ F^k(\btheta^{(i)}, \momentum^{(i)}), k = - L^{(i)}, \ldots, K - L^{(i)} \}
		\end{equation}
		Sample $(\btheta_k^{(i+1)}, \momentum_k^{(i+1)})$'s by independently setting $(\btheta_k^{(i+1)}, \momentum_k^{(i+1)}) = (\btheta^*, \momentum^*) \in \mathcal{A}_{i+1}$ with probability
		\begin{equation}
		\label{eq:bernton_weight}
		w_{i+1}(\btheta^*, \momentum^*)
		= \frac{\pi(\btheta^*, \momentum^*)}{\sum_{(\btheta, \momentum) \in \mathcal{A}_{i+1}} \pi(\btheta, \momentum) }.
		\end{equation}
	\end{enumerate}
\end{algo}

Taking an expectation over the sampling procedure of $\{ (\btheta_k^{(i+1)}, \momentum_k^{(i+1)}), k = 1, \ldots, K\}$ in Step~4 above, we obtain the Rao-Blackwellized version of Algorithm~\ref{alg:calderhead}.
\begin{algo}[Rao-Blackwellization of Algorithm~\ref{alg:calderhead}]
	\label{alg:bernton}
	Given the collection of states $\A_i$ with the weights $w_i$ as in \eqref{eq:calderhead_proposal_states} and \eqref{eq:bernton_weight}, return the weighted empirical measure 
	\begin{equation*}
	\frac{1}{N} \sum_{i=1}^N \sum_{(\btheta^*, \momentum^*) \in \A_i} w_i(\btheta^*, \momentum^*) \delta_{(\btheta^*, \momentum^*)} (\cdot)
	\end{equation*}
	as a Monte Carlo estimate of the target distribution.
\end{algo}

\begin{proof}[Proof of Validity of Algorithm~\ref{alg:calderhead}]
	We will establish the weak convergence 
	\begin{equation}
	\frac{1}{N K} \sum_{i=1}^N \sum_{k =1}^{K} \delta_{(\btheta_k^{(i)}, \momentum_k^{(i)})} (\cdot) \overset{w}{\to} \pi(\cdot)
	\text{ as } N \to \infty.
	\end{equation}
	for the samples $\{ (\btheta_k^{(i)}, \momentum_k^{(i)}), k = 1, \ldots, K\}$ generated as in Algorithm~\ref{alg:calderhead}. 
	
	Let $P_0(\cdot \given \cdot)$ denote the transition kernel corresponding to the transition rule $(\btheta_0^{(i)}, \momentum_0^{(i)}) \to (\btheta_0^{(i+1)}, \momentum_0^{(i+1)})$ as in Step 1--3 of Algorithm~\ref{alg:calderhead}. Also let $P_1(\cdot \given \cdot) = \ldots = P_K(\cdot \given \cdot)$ denote the kernel corresponding to the transition rule $(\btheta_0^{(i)}, \momentum_0^{(i)}) \to (\btheta_1^{(i+1)}, \momentum_1^{(i+1)})$ as in Step 4 of Algorithm~\ref{alg:calderhead}. By virtue of Theorem~\ref{thm:vector_ergodic_theorem}, we simply need to verify that $\pi(\cdot)$ is the stationary distribution of the transition kernels $P_0(\cdot \given \cdot)$ and $P_1(\cdot \given \cdot)$. The kernel $P_0(\cdot \given \cdot)$ represents the transition rule of HMC with a randomized number of leap-frog steps and hence is reversible with respect to $\pi(\cdot)$. The reversibility of $P_1(\cdot \given \cdot)$ also follows from the standard HMC theory; the only additional observation needed for the proof is the following ``symmetry'' in the collection of proposed states at Step~4. For $L \sim {\rm Uniform}(\{0, 1, \ldots, K\})$ and a pair of states $(\btheta, \momentum)$ and $(\btheta^*, \momentum^*)$, the following conditional distributions of random sets $\A(\btheta, \momentum, L)$ and $\A(\btheta^*, \momentum^*, L)$ as defined in \eqref{eq:calderhead_proposal_states} coincide:
	\begin{equation*}
	\A(\btheta, \momentum, L) \given (\btheta^*, \momentum^*) \in \A(\btheta, \momentum, L) 
	\overset{d}{=} \A(\btheta^*, \momentum^*, L) \given (\btheta, \momentum) \in \A(\btheta^*, \momentum^*, L) 
	\qedhere
	\end{equation*}
\end{proof}

\section{Efficient Recycled NUTS}
\label{app:efficient_recycled_NUTS}
Rao-Blackwellized recycled NUTS of Algorithm~\ref{alg:rao_blackwell_ReNUTS} is statistically efficient, but requires all the intermediate states. As an alternative, we here describe a modification of Algorithm~\ref{alg:simple_ReNUTS} which improves statistical efficiency without increasing the number of recycled states by ensuring that the recycled samples are evenly spread across a NUTS trajectory $\T(\btheta_0^{(i-1)}, \momentum_0^{(i-1)})$ . As mentioned in Section~\ref{sec:recycled_NUTS}, we can take advantage of the binary tree structure of $\T(\btheta_0^{(i-1)}, \momentum_0^{(i-1)})$ in order to implement it in a simple and memory efficient manner. To explain the main idea, suppose that an iteration of NUTS from $(\btheta_0^{(i-1)}, \momentum_0^{(i-1)})$ takes $2^d - 1$ leapfrog steps, generating a binary tree of depth $d$ which we denote by $\T_d$. Let $\T_{d-j, \ell}$ and $\A_{d-j, \ell} = \A(\T_{d-j, \ell})$ for $\ell = 1, \ldots, 2^{j}$ denote the subtrees of depth $d - j$ and the collections of acceptable states within each subtree. At the depth $d-j$, we assign the minimal number of samples from the subtree $\T_{d-j, \ell}$ to be
\begin{equation*}
\left\lfloor K \frac{|\A_{d-j, \ell}|}{\sum_\ell |\A_{d-j, \ell}|} \right\rfloor
\end{equation*}
where $\lfloor \cdot \rfloor$ denotes a floor function. Enforcing this recursively at each depth $d - j$ for $j = 1, \ldots, d$ ensure that the recycled states are evenly spread along the trajectory. An actual procedure is described in Algorithm~\ref{alg:ReNUTS} below. In order to avoid references to the algorithm implementation details of NUTS, we describe how to carry out the recycling procedure assuming we store all the intermediate states and its binary tree structure during each NUTS iteration. It is however easy to add the recycling algorithm on top of the NUTS implementation of \cite{hoffman14} so that no more than $K$ states are stored in memory during each NUTS iteration. 

\begin{algo}[Recycled NUTS]
	\label{alg:ReNUTS}
	Run NUTS to generate a sequence of random variables $\{ (\btheta_0^{(i)}, \momentum_0^{(i)}) \}_{i \geq 1}$. Additionally at each iteration of NUTS, recycle variables $\{ (\btheta_k^{(i)}, \momentum_k^{(i)}), k = 1, \ldots, K \}$ from the collection of acceptable states $\A(\btheta_0^{(i-1)}, \momentum_0^{(i-1)})$ by calling the function \textsc{Recycle} below: 
	
	\begin{minipage}{.9\textwidth}
		\vspace{2pt}
		\rule{\textwidth}{.3pt}
		\begin{algorithmic}[]
			\setlength{\leftskip}{-1.5em}
			\Function{Recycle}{$\A$, $K$}
			\If{${\rm depth}(\A) = 0$} \Comment{\textit{$A$ is a singleton set}}
			\State \Return $K$ copies of the variable from $\A$
			\Else
			\State let $\A'$ and $\A''$ be the left and right subtree of $\A$
			\State $n \leftarrow \lfloor w \rfloor + {\rm Bernoulli}(w - \lfloor w \rfloor) $ \ for $w = K / |\A'|$
			\State %$\{(\btheta_1, \momentum_1), \ldots,(\btheta_n, \momentum_n) \} \leftarrow {\rm Recycle}(\A', n)$
			$\{(\btheta_k, \momentum_k)\}_{k=1}^{n} \leftarrow {\rm Recycle}(\A', n)$
			\State %\mbox{$\{(\btheta_{n+1}, \momentum_{n+1}), \ldots,(\btheta_K, \momentum_K)\} \leftarrow {\rm Recycle}(\A'', K - n)$}
			$\{(\btheta_k, \momentum_k)\}_{k=n+1}^{K} \leftarrow {\rm Recycle}(\A'', K - n)$
			\State \Return $\{(\btheta_1, \momentum_1), \ldots,(\btheta_K, \momentum_K) \}$
			\EndIf
			\EndFunction
		\end{algorithmic}
		\vspace{-2ex} 
		\rule{\textwidth}{.3pt}
	\end{minipage}
\end{algo}

\section{ESS gain as a function of chain lengths}
\label{app:ess_ratio_as_function_of_chain_length}
\begin{figure}[htb]
	\centering
	\begin{subfigure}{0.45\textwidth}
		\includegraphics[width=\textwidth]{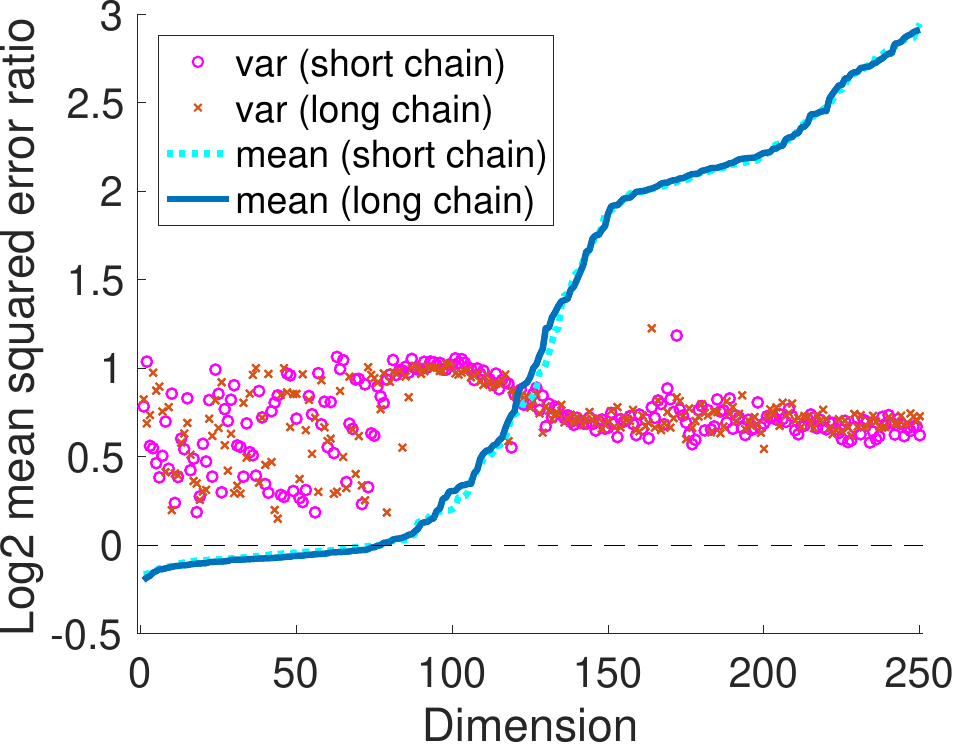} 
	\end{subfigure}
	~
	\begin{subfigure}{0.45\textwidth}
		\includegraphics[width=.96\textwidth]{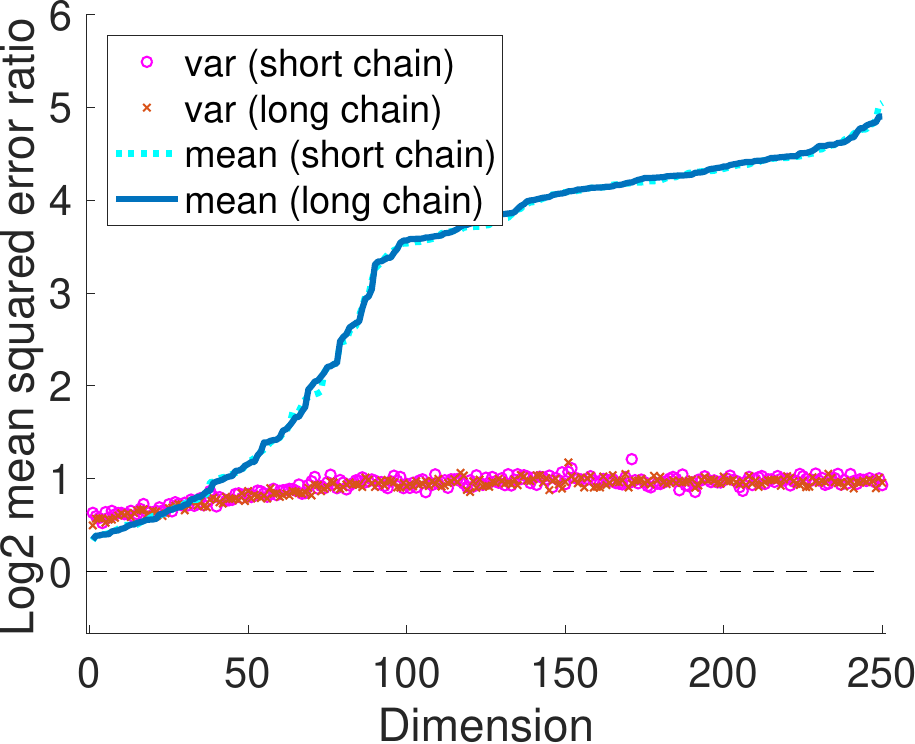}
	\end{subfigure}
	\caption{Comparison of the $\log_2$ ESS ratios under the short and long chain for the multivariate Gaussian example.}
	\label{fig:Gaussian_chain_length_comparison}
\end{figure}

\begin{figure}[htb]
	\centering
	\begin{subfigure}{0.45\textwidth}
		\includegraphics[width=\textwidth]{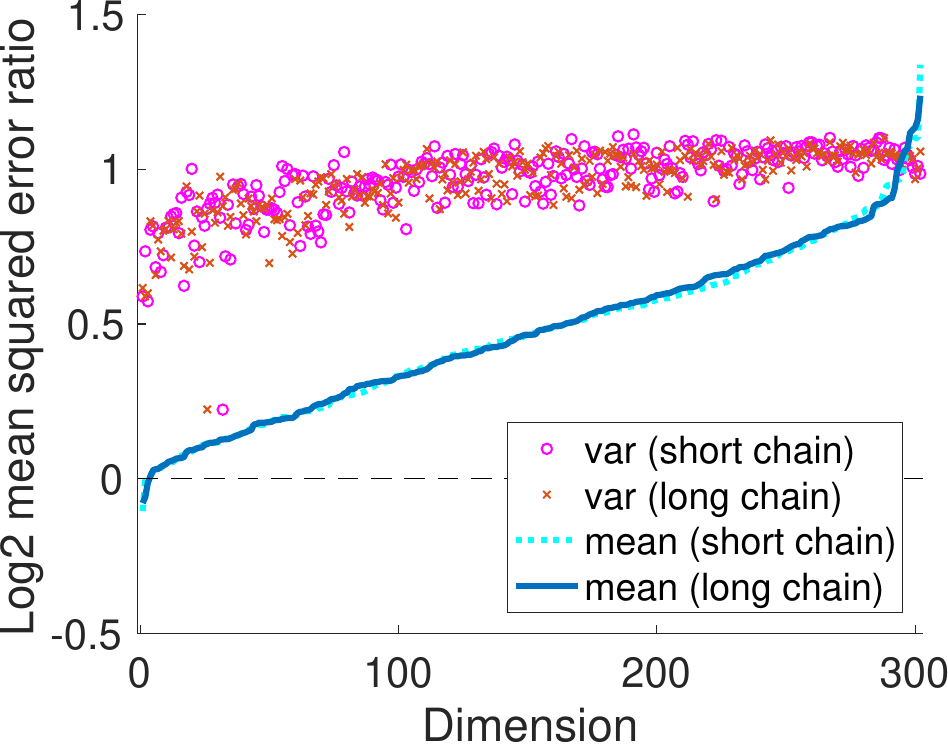} 
	\end{subfigure}
	~
	\begin{subfigure}{0.45\textwidth}
		\includegraphics[width=.96\textwidth]{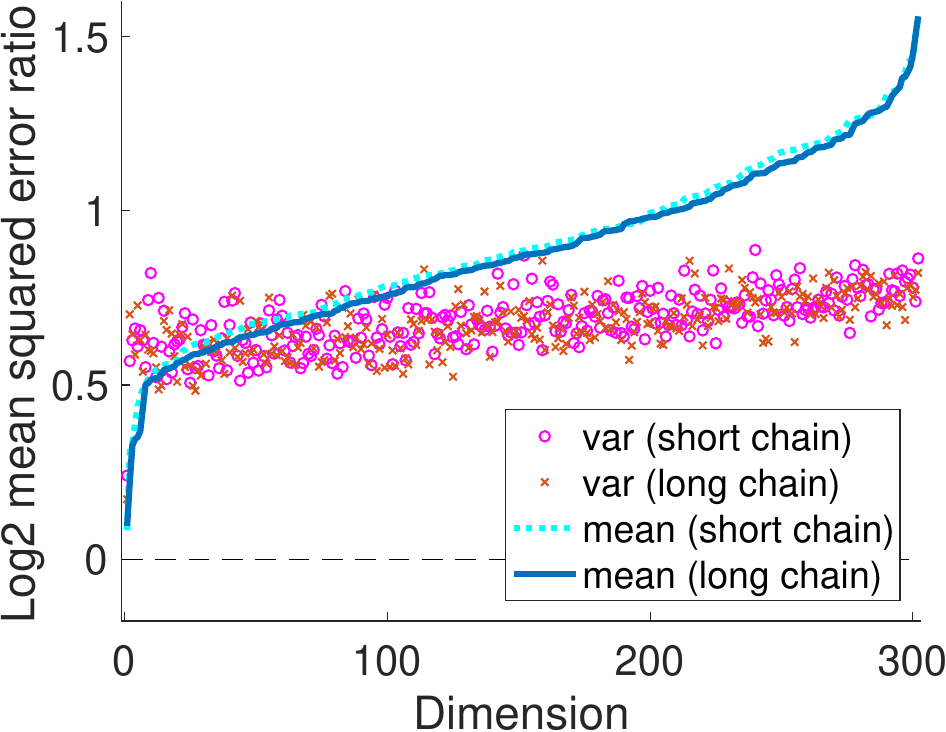}
	\end{subfigure}
	\caption{Comparison of the $\log_2$ ESS ratios under the short and long chain for the hierarchical Bayesian logistic regression example.}
	\label{fig:HBLR_chain_length_comparison}
\end{figure}

\begin{figure}[htb]
	\centering
	\begin{subfigure}{0.45\textwidth}
		\includegraphics[width=\textwidth]{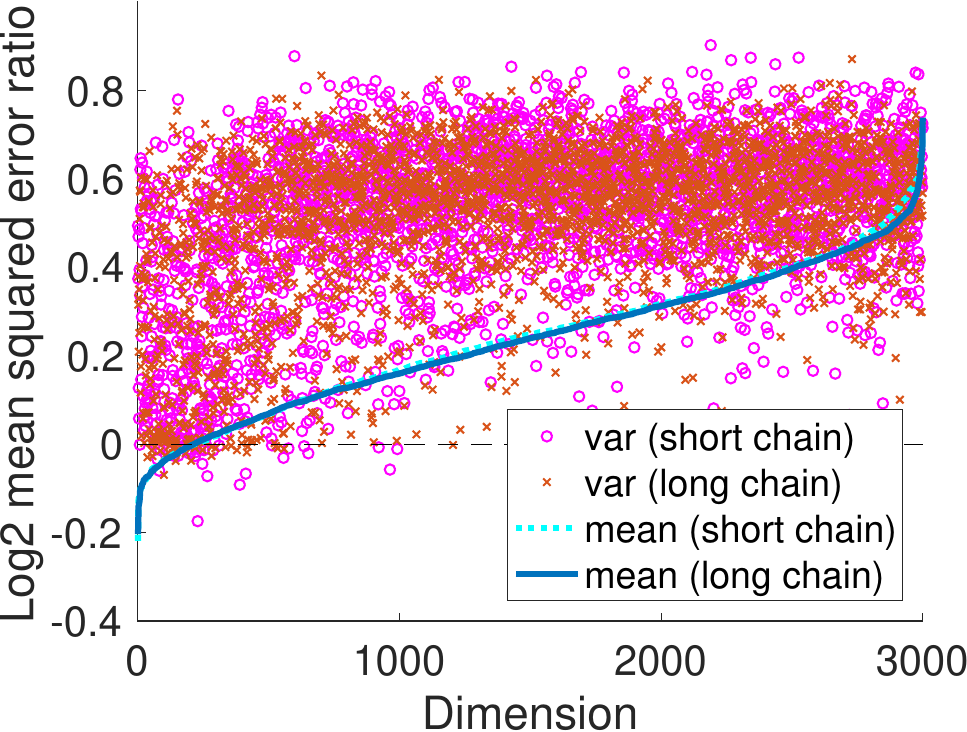} 
	\end{subfigure}
	~
	\begin{subfigure}{0.45\textwidth}
		\includegraphics[width=.96\textwidth]{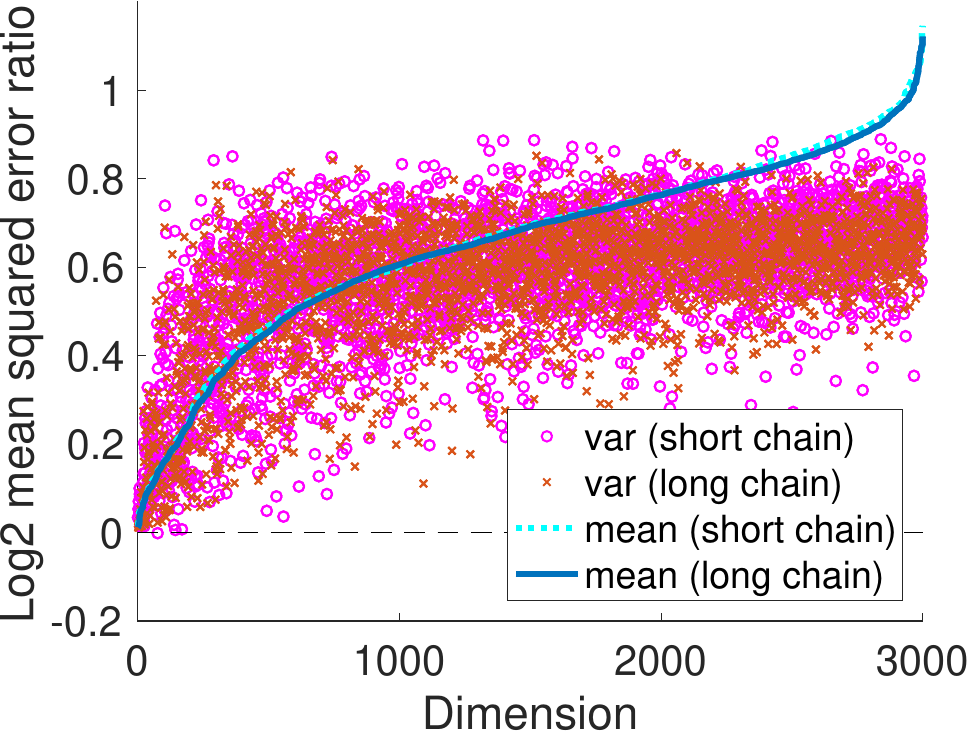}
	\end{subfigure}
	\caption{Comparison of the $\log_2$ ESS ratios under the short and long chain for the stochastic volatility model example.}
	\label{fig:SV_chain_length_comparison}
\end{figure}

\begin{figure}[htb]
	\centering
	\begin{subfigure}{0.45\textwidth}
		\includegraphics[width=\textwidth]{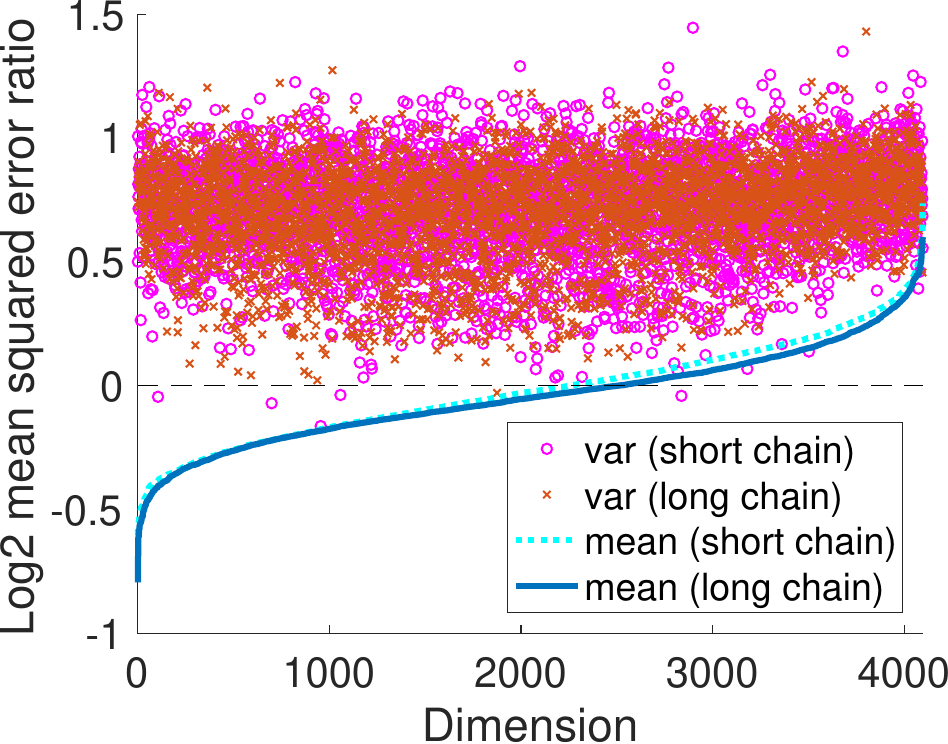} 
	\end{subfigure}
	~
	\begin{subfigure}{0.45\textwidth}
		\includegraphics[width=.96\textwidth]{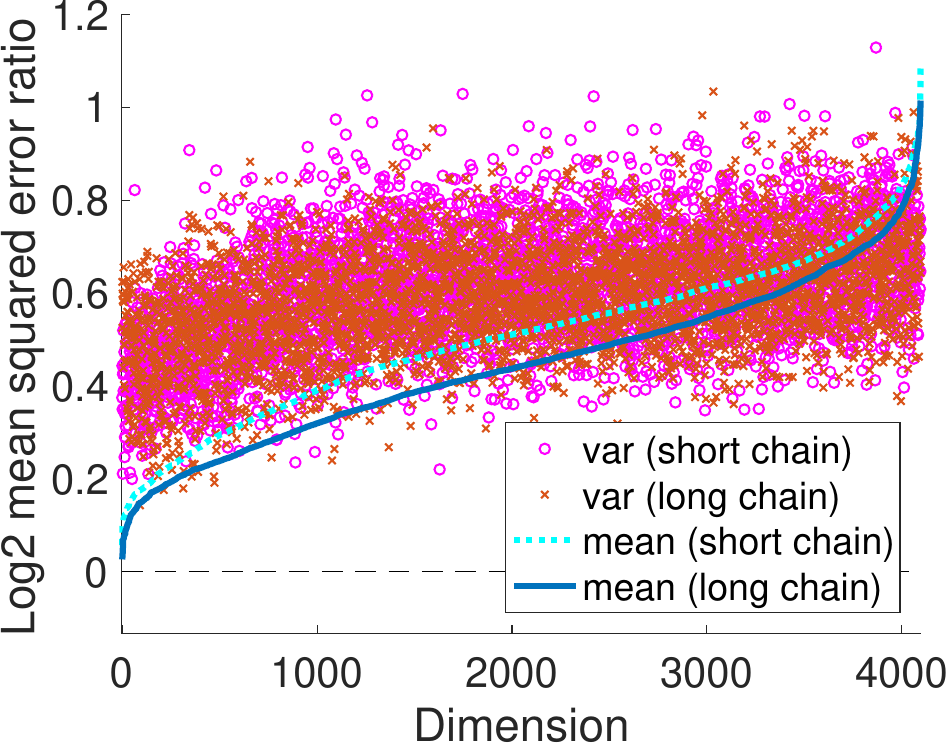}
	\end{subfigure}
	\caption{Comparison of the $\log_2$ ESS ratios under the short and long chain for the log-Gaussian Cox model.}
	\label{fig:logGaussCox_chain_length_comparison}
\end{figure}

Here we investigate how the ESS gain from recycling depends on the chain length for the first moment, second moment, and quantile estimators considered in Section~\ref{sec:simulation}. To this end, we record the ESS ratios at 1/32, 1/16, 1/8, 1/4, and 1/2 of the total chain length (1,600 for the log-Gaussian Cox example and 3,200 for all others). For all the three estimators, we find that the ESS ratios across the chain lengths. We summarize the findings by plotting the $\log_2$ ESS ratios as in Section~\ref{sec:simulation} for both the chain of the full length and of 1/32th length. These are shown in Figure~\ref{fig:Gaussian_chain_length_comparison} -- \ref{fig:logGaussCox_chain_length_comparison}. We show the results only for the first and second moment estimators to avoid cluttering the plots, but the results for the quantile estimator demonstrate essentially the identical conclusion. 

\FloatBarrier

% ** Acknowledgements **
% \begin{acknowledgement}
% \end{acknowledgement}

\end{document}